\documentclass[%
 twocolumn,
 superscriptaddress,
 nofootinbib,
 showpacs,preprintnumbers,
 aps,
 pra
]{revtex4-1}

\usepackage{braket}
\usepackage{graphicx}
\usepackage{dcolumn}
\usepackage{bm}
\usepackage{newtxtext,newtxmath}
\usepackage{amsmath,amssymb}

\newcommand{\Tr}{\mathrm{Tr}}
\newcommand{\bes}{\begin{subequations}}
\newcommand{\ees}{\end{subequations}}
\usepackage[up]{subfigure}
\graphicspath{{fig/}}
\begin{document}

\title{Quantum annealing of the $p$-spin model under inhomogeneous transverse field driving}

\author{Yuki Susa}
\thanks{Present address: System Platform Research Laboratories, NEC\\ Corporation, 34 Miyukigaoka, Tsukuba, Ibaraki 305-8501, Japan;\\ y-susa@bx.jp.nec.com}
\affiliation{Institute of Innovative Research, Tokyo Institute of Technology, Oh-okayama, Meguro-ku, Tokyo 152-8550, Japan}

\author{Yu Yamashiro}
\affiliation{Department of Physics, Tokyo Institute of Technology, Oh-okayama, Meguro-ku, Tokyo 152-8550, Japan}

\author{Masayuki Yamamoto}
\affiliation{Graduate School of Information Sciences, Tohoku University, Sendai 980-8579, Japan}

\author{Itay Hen}
\affiliation{Center for Quantum Information Science \& Technology, University of Southern California, Los Angeles, California 90089, USA}
\affiliation{Department of Physics \& Astronomy, University of Southern California, Los Angeles, California 90089, USA}
\affiliation{Information Sciences Institute, University of Southern California, Marina del Rey, California 90292, USA}

\author{Daniel A. Lidar}
\affiliation{Center for Quantum Information Science \&
Technology, University of Southern California, Los Angeles, California 90089, USA}
\affiliation{Department of Physics \& Astronomy, University of Southern California, Los Angeles, California 90089, USA}
\affiliation{Department of Electrical Engineering, University of Southern California, Los Angeles, California 90089, USA}
\affiliation{Department of Chemistry, University of Southern California, Los Angeles, California 90089, USA}

\author{Hidetoshi Nishimori}
\affiliation{Institute of Innovative Research, Tokyo Institute of Technology, Oh-okayama, Meguro-ku, Tokyo 152-8550, Japan}

\date{\today}
\begin{abstract}
We solve the mean-field-like $p$-spin Ising model under a spatiotemporal inhomogeneous transverse field to study the effects of inhomogeneity on the performance of quantum annealing. We previously found that the problematic first-order quantum phase transition that arises under the conventional homogeneous field protocol can be avoided if the temperature is zero and the local field is completely turned off site by site after a finite time. We show in the present paper that, when these ideal conditions are not satisfied, another series of first-order transitions appear, which prevents us from driving the system while avoiding first-order transitions.  Nevertheless, under these nonideal conditions, quantitative improvements can be obtained in terms of narrower tunneling barriers in the free-energy landscape. A comparison with classical simulated annealing establishes a limited quantum advantage in the ideal case, since inhomogeneous temperature driving in simulated annealing cannot remove a first-order transition, in contrast to the quantum case. The classical model of spin-vector Monte Carlo is also analyzed, and we find it to have the same thermodynamic phase diagram as the quantum model in the ideal case, with deviations arising at non-zero temperature.
\end{abstract}

\maketitle
\section{Introduction}
Quantum annealing (QA) is a metaheuristic for combinatorial optimization problems and is closely related to adiabatic quantum computation \cite{kadowaki1998quantum,farhi2001quantum,santoro2002theory,santoro2006optimization,das2008colloquium,Morita2008,albash2018adiabatic}, in which the final-time classical ground state of an Ising Hamiltonian encodes the optimal solution of a combinatorial optimization problem \cite{Lucas2014}. Quantum fluctuations are applied to the Ising model, first with a very large amplitude and then slowly reduced to zero, to reach the ground state of the original Ising model representing the solution to the combinatorial optimization problem. The amplitude of quantum fluctuations is a key control parameter, analog to the temperature in the classical analog, simulated annealing \cite{Kirkpatrick1983}.

As the amplitude of quantum fluctuations is reduced, quite generally a quantum phase transition takes place in the thermodynamic limit and at zero temperature from a disordered paramagnetic phase to an ordered phase. The existence of such a phase transition can be a serious problem for QA because it may slow down the annealing process significantly. This can be understood in terms of the adiabatic theorem of quantum mechanics, which states that a sufficient condition for the system to stay in the instantaneous ground state is that the total evolution time is inversely proportional to a polynomial of the energy gap between the instantaneous ground state and the first excited state \cite{Jansen:07,lidar:102106}. It is known empirically that the energy gap decreases exponentially as a function of the system size at a first-order quantum phase transition\footnote{A few exceptions exist as exemplified, e.g., in Refs.~\cite{Laumann:2012hs,Tsuda2013}.} whereas the scaling of gap decrease is significantly milder, i.e., polynomial in the system size, at a second-order transition as expected generally from finite-size scaling \cite{nishimori2011phasetransition}.  This, in combination with the adiabatic theorem, means that the order of a quantum phase transition, or its mere existence, can be a decisive factor for the efficiency of QA in its adiabatic realization, because the time complexity grows exponentially for a first-order transition but is polynomial at a second-order transition or for the case of no transition.\footnote{The leading contribution to the computation time or computational complexity is the denominator (the energy gap) of the formula for the adiabatic theorem. However, when the gap stays finite as in the case without transition, the numerator dominates the behavior, which is polynomial in system size.} The situation is considerably more complicated at finite temperature in an open system, where the quantum adiabatic theorem involves the gap of the Liouvillian rather than the Hamiltonian \cite{Avron:2012tv,Venuti:2015kq}. Nevertheless, similar scaling considerations apply \cite{Venuti:2017aa}.

While the phase-transition perspective is certainly not sufficient for a complete understanding of the scaling of QA-based algorithms, since there does not exist a strict relation between the static properties in the thermodynamic limit and the dynamic properties at finite system size, it is nevertheless an insightful heuristic amenable to an analytical treatment that allows one to anticipate the finite-size scaling behavior, and we adopt it here for this reason, in line with a recent series of other studies, e.g., Refs.~\cite{Jorg:2008aa,jorg2010energy,Nishimori:2015dp,MNAL:15,Matsuura:2016aa}.\footnote{We nevertheless should keep in mind that there exist examples in which thermodynamic calculations do not necessarily lead to the correct understanding of finite-size properties of quantum systems~\cite{Laumann:2015sw,Durkin:2018aa}.} In the same vein, efforts have been invested to reduce the difficulty arising from a first-order transition by, for example, the increase of the order of the transition from first to second using nonstoquastic Hamiltonians \cite{seki2012quantum,seoane2012many,Seki2015,nishimori2017exponential} or by the reverse annealing protocol \cite{Ohkuwa2018}.

Recently, the protocol of inhomogeneous driving of the transverse field has been studied as a candidate to enhance the performance of QA.  In this method, one changes the amplitude of quantum fluctuations site by site individually.\footnote{We use the terms ``site" and ``spin" interchangeably.} For example, the one-dimensional ferromagnetic Ising model with weak disorder was studied in Refs.~\cite{rams2016inhomogeneous,Mohseni:2018aa}, where the residual energy was found to be smaller than in the homogeneous case.  Similar improvements by inhomogeneous driving were reported in one-dimensional models in Refs.~\cite{Dziarmaga2010,Zurek2008}.  Inhomogeneous field driving for the random 3-SAT problem has been shown to mitigate difficulties near the end of annealing processes by numerical computations in Ref.~\cite{Farhi2011}. Avoidance of problematic anticrossings near the end of the anneal was also discussed analytically in Refs.~\cite{Dickson2011,Dickson2012} and was tested on an experimental quantum annealer \cite{Lanting2017}. See also Refs.~\cite{del2013causality,gomez2018universal,Adame:2018aa} for related studies.

Given these circumstances, several of the present authors solved the ferromagnetic $p$-spin model under inhomogeneous driving of the transverse field exactly\footnote{In Ref.~\cite{okuyama2018exact}, it has been shown that the ``static approximation" used in Ref.~\cite{susa2018exponential} leads to the exact solution in the present problem.} and showed that first-order transitions can be removed if the inhomogeneity of the field is appropriately controlled \cite{susa2018exponential}.  However, the analysis in Ref.~\cite{susa2018exponential} is valid under idealized conditions such as the zero-temperature limit and complete turning off of the field at each site after a finite amount of time. Here we generalize this previous study and investigate what happens under more realistic conditions, including a nonzero temperature.  We also compare the quantum system with its classical counterparts to clarify if and how quantum effects are essential in the present problem.

This paper is organized as follows. In Sec. \ref{sec:formulation}, we formulate the problem. In Sec. \ref{sec:inhomogeneous_QA}, we examine the effects of inhomogeneous driving of the transverse field under idealized conditions. Section \ref{sec:practical_situation} removes some of those conditions. In Sec. \ref{sec:classical}, we consider two classical approaches, simulated annealing with site dependent temperature and the spin-vector Monte Carlo method. The final section is devoted to conclusions.

\section{Formulation}

\label{sec:formulation}
We write the Hamiltonian of QA as
\begin{align}
\label{eq:total_Hamiltonian}
\hat{H}(s)=s\hat{H}_0+\hat{V},
\end{align}
where $\hat{H}_0$ is the target Hamiltonian, the ground state of which encodes the solution to a given combinatorial optimization problem, $\hat{V}$ is the driver Hamiltonian used to induce quantum fluctuations, and $s$ is a dimensionless parameter that controls the time dependence.  We choose the $p$-spin model as the target Hamiltonian,
\begin{align}
\label{eq:target}
\hat{H}_0=-N\left(\frac{1}{N} \sum_{i=1}^{N}\hat{\sigma}_i^z\right)^{p},
\end{align}
where $p(\ge 3)$ is an integer, $\hat{\sigma}_i^z$ is the $z$ component of the Pauli operator, $N$ is the total number of spins, and $i$ is the site (qubit) index running from $1$ to $N$.

The ground state of $\hat{H}_0$ is trivial, $\otimes_{i=1}^N \ket{0}_i$ for odd $p$, where $\ket{0}_i$ denotes the spin-up state, i.e., $\hat{\sigma}_i^z\ket{0}_i=\ket{0}_i$. For even $p$, another state $\otimes_{i=1}^N \ket{1}_i$ is also a ground state, where  $\hat{\sigma}_i^z\ket{1}_i=-\ket{1}_i$.   This model reduces to the Grover problem \cite{Grover:97a} in the limit $p\to\infty$ \cite{jorg2010energy}.

We choose the driver Hamiltonian in the following form: 
\begin{align}
\label{eq:driver_Hamiltonian}
\hat{V}=-\sum_{i=1}^{N} \Gamma_i \hat{\sigma}_i^x,
\end{align} 
where $\hat{\sigma}_i^x$ is the $x$ component of the Pauli operator.  We assume $\Gamma_i\ge 0$ without loss of generality. 

Let us briefly recall the situation under conventional QA, where the coefficient $\Gamma_i$ satisfies $\Gamma_i=1-s$, which is homogeneous in $i$. In this case the ground state of the driver Hamiltonian is trivial, $\otimes_{i=1}^{N}(\ket{0}_i+\ket{1}_i)/\sqrt{2}$. As time evolves, $s$ increases from $0$ to $1$, and the Hamiltonian (\ref{eq:total_Hamiltonian}) changes from $\hat{V}$ at $s=0$ to $\hat{H}_0$ at $s=1$.  Under this homogeneous transverse field, it is known that QA for the $p$-spin model has a first-order phase transition for $p\ge 3$ \cite{jorg2010energy}. This would appear to be a disturbing failure of QA, since the optimization problem is trivial but is difficult for QA, although classical simulated annealing also fails due to a first-order thermal phase transition. However, it is possible to change this first-order transition to second order by the introduction of antiferromagnetic transverse interactions, which makes the Hamiltonian nonstoquastic \cite{seki2012quantum,seoane2012many,Seki2015}. It is also possible to remove the transition by reverse annealing \cite{Ohkuwa2018}.

An alternative way to circumvent the difficulties of first-order transitions is via spatiotemporal inhomogeneity of the transverse field \cite{susa2018exponential}:
\begin{align}
\label{eq:inhomogeneous_schedule}
\Gamma_i=
\begin{cases}
1 &\text{for}\ \  0\leq i/N\leq 1-\tau, \\
N(1-\tau)+(1-i) &\text{for}\ \  1-\tau<i/N<1-\tau+1/N, \\
0 &\text{for}\ \  1-\tau+1/N\leq i/N\leq1.
\end{cases}
\end{align}
Here, $\tau$ is another dimensionless time-dependent parameter varying from $0$ to $1$, used to control the number of spins under the influence of the transverse field. This describes a step function with a diagonal drop. In the limit $N\gg 1$, the drop becomes vertical [the range of $i$ in the middle line on the right-hand side of Eq.~(\ref{eq:inhomogeneous_schedule}) becomes negligible] and the following form is asymptotically correct:
\begin{align}
\Gamma_i=
\begin{cases}
1 &\text{for}\ \  0\leq i/N\leq 1-\tau, \\
0 &\text{for}\ \  1-\tau < i/N\leq1.
\end{cases}
\label{eq:inhomogeneous_schedule2}
\end{align}
In this limit the driver Hamiltonian $\hat{V}$ with the above $\Gamma_i$ reduces to the simple form
\begin{align}
    \hat{V}=-\sum_{i=1}^{N(1-\tau)} \hat{\sigma}_i^x ,
    \label{eq:simple_TF}
\end{align}
which describes a ``zipper-closing"-like schedule for the transverse field, starting from the last site.

\section{Idealized case}
\label{sec:inhomogeneous_QA}
We first recapitulate the idealized case with the transverse field applied only to a part of the system as in Eq.~(\ref{eq:simple_TF}) at zero temperature as studied in Ref.~\cite{susa2018exponential}. We can derive an explicit form of the free energy for the Hamiltonian (\ref{eq:total_Hamiltonian}) with the $p$-spin model (\ref{eq:target}) and the general driver Hamiltonian (\ref{eq:driver_Hamiltonian}) by the standard method of the Suzuki-Trotter decomposition in combination with the static approximation. We delegate the details to Appendix \ref{app:A} and just write the results for the free energy per spin and the self-consistent equation for the magnetization at finite temperature $T(=1/\beta)$:
\bes
\begin{align}
\label{eq:free_energy_finite_temp}
f(m)=&s(p-1) m^p \notag \\
&-\frac{1}{\beta} \int_0^1 dx \ln 2\cosh \beta \sqrt{(spm^{p-1})^2+\Gamma(x)^2}, \\
m=&\int_0^1 dx \frac{spm^{p-1}}{\sqrt{(spm^{p-1})^2+\Gamma(x)}} \notag \\
&\times\tanh \beta \sqrt{(spm^{p-1})^2+\Gamma(x)},
\label{eq:SCE1}
\end{align}
\ees
respectively, where $x$ is the normalized site index $i/N$ in the continuous (large-$N$) limit. In the zero-temperature limit, these equations reduce to
\bes
\begin{align}
\label{eq:free_energy_zero_temp}
f(m)=&s(p-1) m^p -\int_0^1 dx  \sqrt{(spm^{p-1})^2+\Gamma(x)^2}, \\
m=&\int_0^1 dx \frac{spm^{p-1}}{\sqrt{(spm^{p-1})^2+\Gamma(x)}}.
\end{align}
\ees
Substituting the continuum limit of 
Eq.~\eqref{eq:inhomogeneous_schedule2} into the free energy (\ref{eq:free_energy_zero_temp}), we reproduce Eq.~(1) of Ref.~\cite{susa2018exponential},
\begin{align}
\label{eq:free_energy}
f(m)=&s(p-1) m^p -(1-\tau) \sqrt{(spm^{p-1})^2+1} \notag \\
&-\tau (spm^{p-1}).
\end{align}

\begin{figure}
  \centering
  \includegraphics[width=0.7\linewidth]{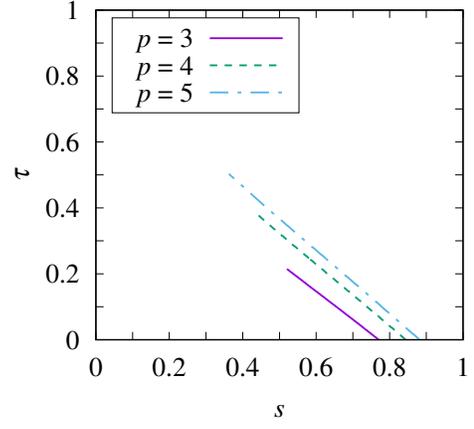}
  \caption{Phase diagram on the $s$-$\tau$ plane for the idealized case at zero temperature. Each color denotes a line of first-order transitions for a given $p$, which is chosen to be 3, 4, and 5.}
   \label{fig:phase_diagram}
\end{figure}

We can draw the phase diagram from these equations as in Fig. \ref{fig:phase_diagram}.
The process of annealing starts at $s=\tau=0$ and terminates at $s=\tau=1$. It is seen that we can choose a path that avoids phase transitions between the starting and the ending points. This is to be contrasted with the case of a homogeneous transverse field, corresponding to the $\tau=0$ axis, in which there is no way to avoid a first-order transition.

Another quantity that it would be instructive to look at is the entanglement entropy, which also exhibits the characteristic behavior of phase transitions (or their absence) depending on the path connecting the starting and end points, as described in Appendix \ref{app:semiclassical}.

We can evaluate the energy gap $\Delta$ in the limit of large system size $N\to\infty$ by the standard semiclassical method \cite{seoane2012many,filippone2011quantum} as explained in some detail in Appendix~\ref{app:semiclassical}. The result is
\bes
\begin{align}
    \Delta &={\rm min}(\Delta_{a_1},\Delta_{b}) \\
    \Delta_{a_1} &= \delta\sqrt{1-\epsilon^2}, ~
    \Delta_b = 2sp \{\tau+(1-\tau)\cos \theta_0 \}^{p-1},
\end{align}
\ees
where
\bes
\begin{align}
    \theta_0&=\arg \min_{\theta}\left\{-s[\tau+(1-\tau)\cos \theta]^p-(1-\tau)\sin \theta\right\}\\
    \epsilon&= -\frac{2 \gamma}{\delta}, \\
    \gamma &= -\frac{1}{2}sp(p-1)(1-\tau) \sin^2 \theta_0 \{\tau+(1-\tau)\cos\theta_0 \}^{p-2}, \\
    \delta &= \Delta_b \cos\theta_0+2\sin \theta_0 +2\gamma .
\end{align}
\ees
\begin{figure*}
  \centering
  \subfigure[\ ]{\includegraphics[width=0.32\linewidth]{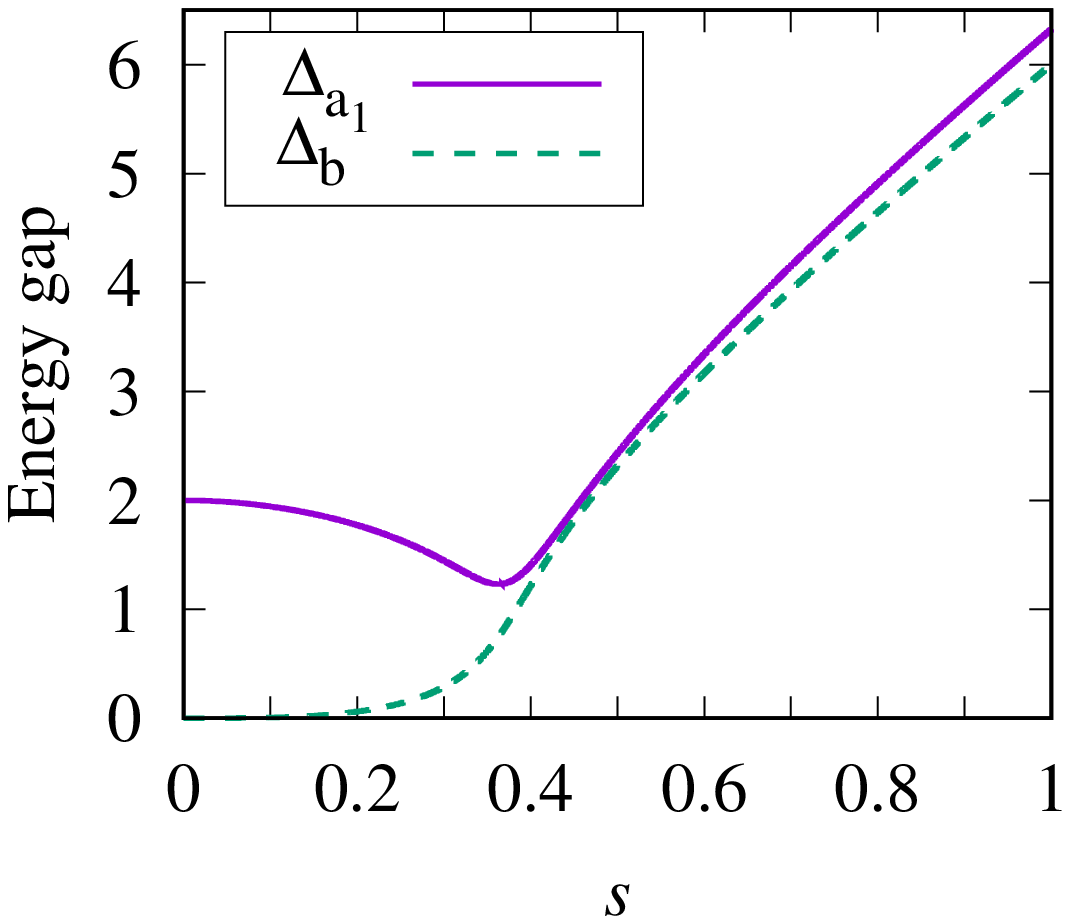}\label{fig:energy_gap-a}} 
  \subfigure[\ ]{\includegraphics[width=0.32\linewidth]{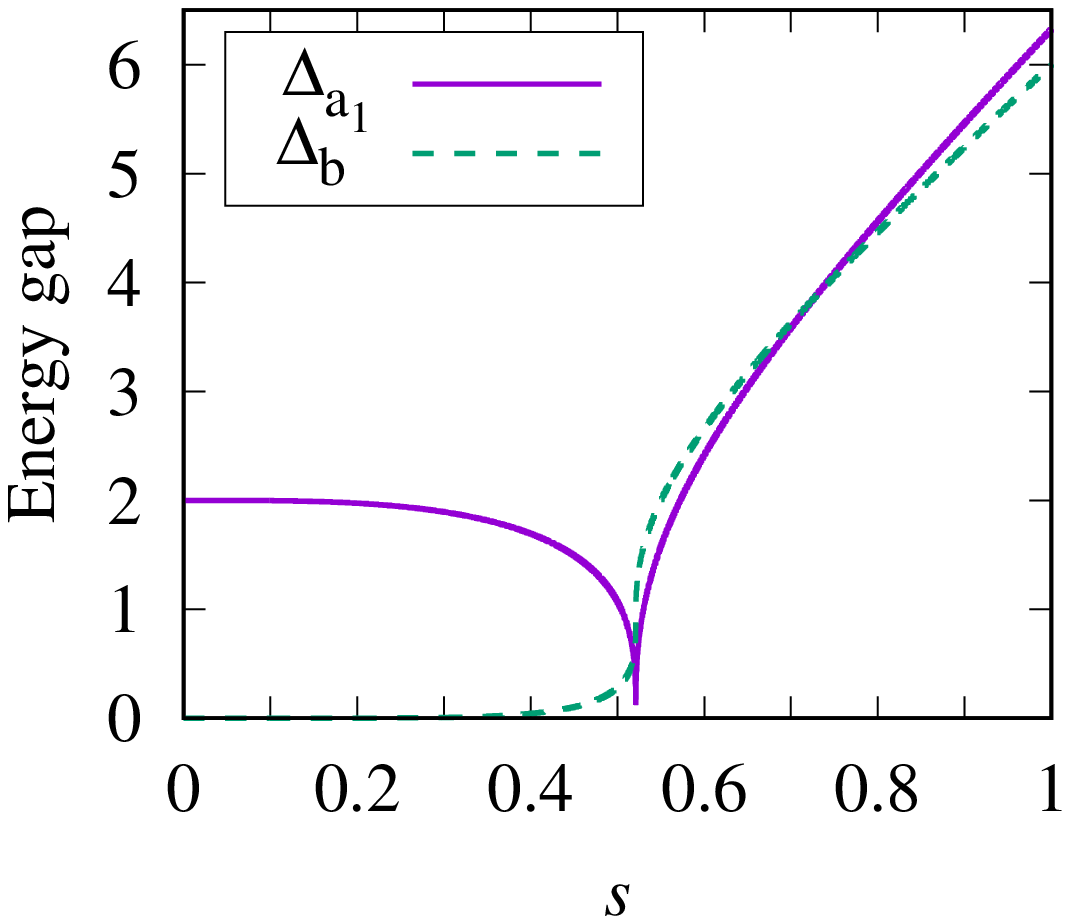}\label{fig:energy_gap-b}} 
  \subfigure[\ ]{\includegraphics[width=0.32\linewidth]{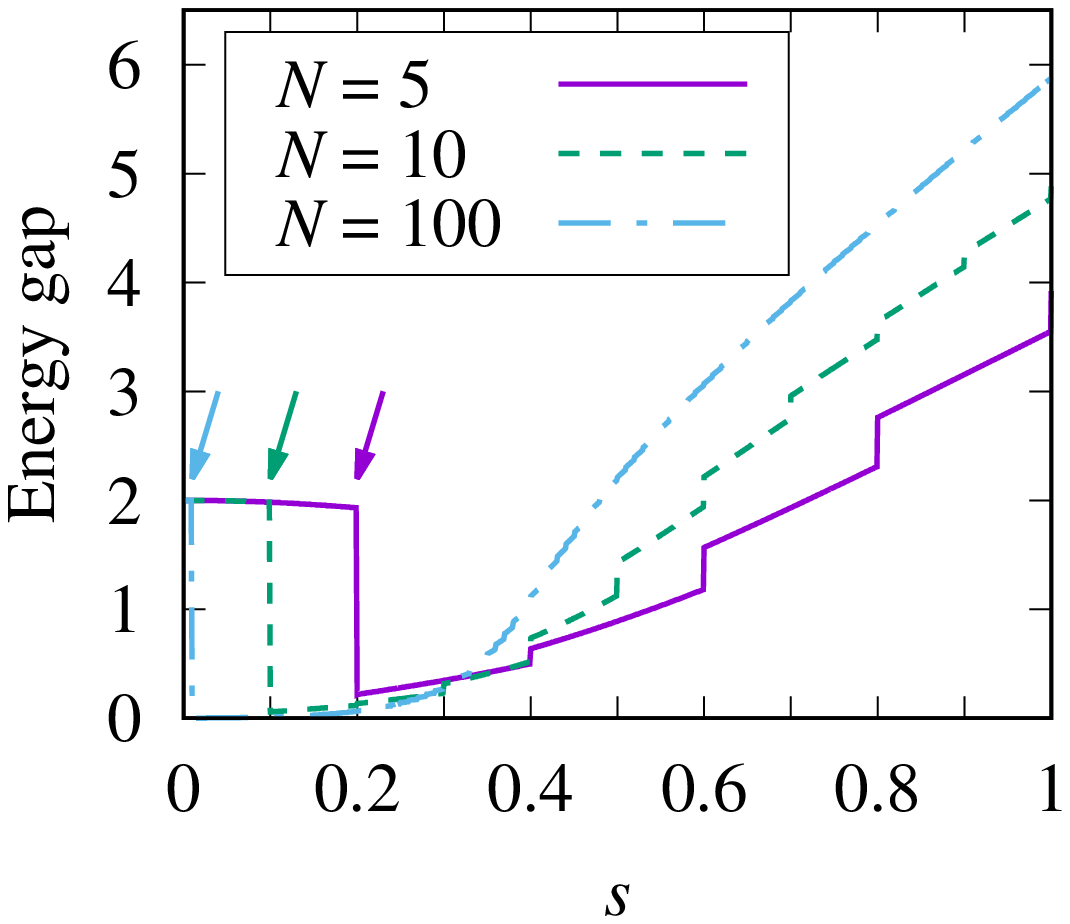}\label{fig:energy_gap-c}}
  \caption{Two types of energy gap $\Delta_{a_1}$ and $\Delta_{b}$ for $p=3$ as functions $s$ for (a) $\tau = s$ (away from the transition line) and (b) $\tau = s^{2.366}$ (just touching the critical point).  The smaller of these two is the final energy gap. (c) The energy gap for finite-size systems with $\tau=s$ obtained by direct numerical diagonalization.  
  The location of the minimum is indicated by an arrow for each $N$.}
   \label{fig:energy_gap}
\end{figure*}
Figures~\ref{fig:energy_gap-a} and \ref{fig:energy_gap-b} show the two energy gap candidates, $\Delta_{a_1}$ and $\Delta_{b}$, for $p=3$ along the paths $\tau = s$, which avoids phase transitions, and $\tau = s^{2.366}$, which just touches the critical point where the first-order line terminates (the paths are illustrated in Fig.~\ref{fig:entanglement_entropy} in Appendix~\ref{app:semiclassical}). The smaller of these two candidates is the true energy gap as shown in Appendix \ref{app:semiclassical}. In Fig.~\ref{fig:energy_gap-a}, $\Delta_{b}$ is seen to be the smaller one and  is a monotonically increasing function of $s$. On the other hand, in Fig.~\ref{fig:energy_gap-b}, the energy gap $\Delta_{a_1}$ is seen to vanish at the critical point $s_c\approx0.52$, as expected.
To check these thermodynamic limit predictions, we calculated the energy gap for finite-size systems by direct numerical diagonalization along the $\tau =s$ path. The result is plotted in Fig.~\ref{fig:energy_gap-c}, which is compatible with the asymptotic behavior in the limit $N\to\infty$ as observed in $\Delta_b$ of Fig.~\ref{fig:energy_gap-a}. It is seen in Fig.~\ref{fig:energy_gap-c} that the energy gap takes its minimum value when the transverse field is turned off at the first site as indicated by the arrows, which implies that the minimum of the gap is located at $s=0$ in the $N\to\infty$ limit.

\begin{figure*}
  \centering
    \subfigure[\ ]{\includegraphics[width=0.3\linewidth]{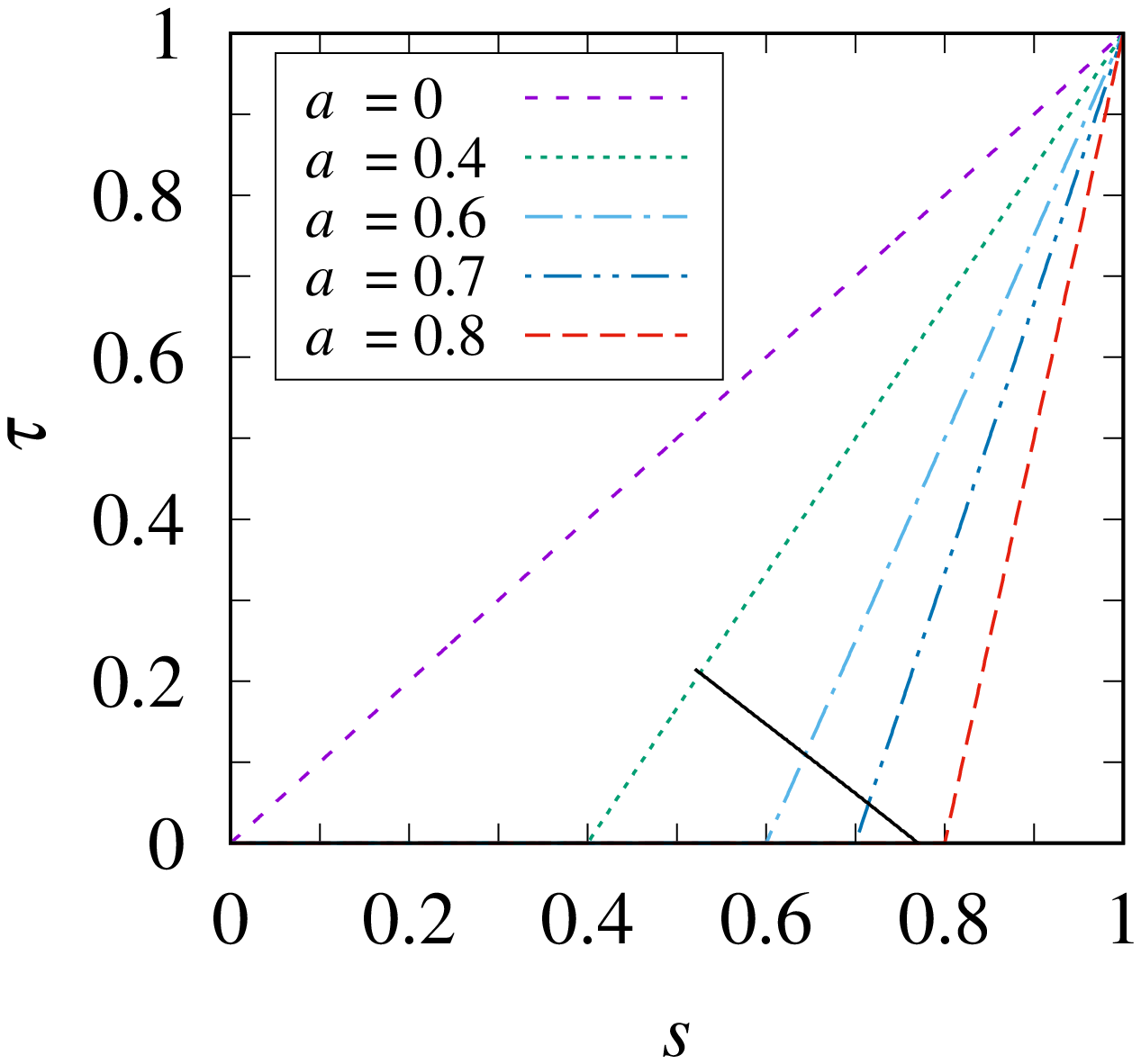}\label{fig:energy_gap2-a}}
    \subfigure[\ ]{\includegraphics[width=0.35\linewidth]{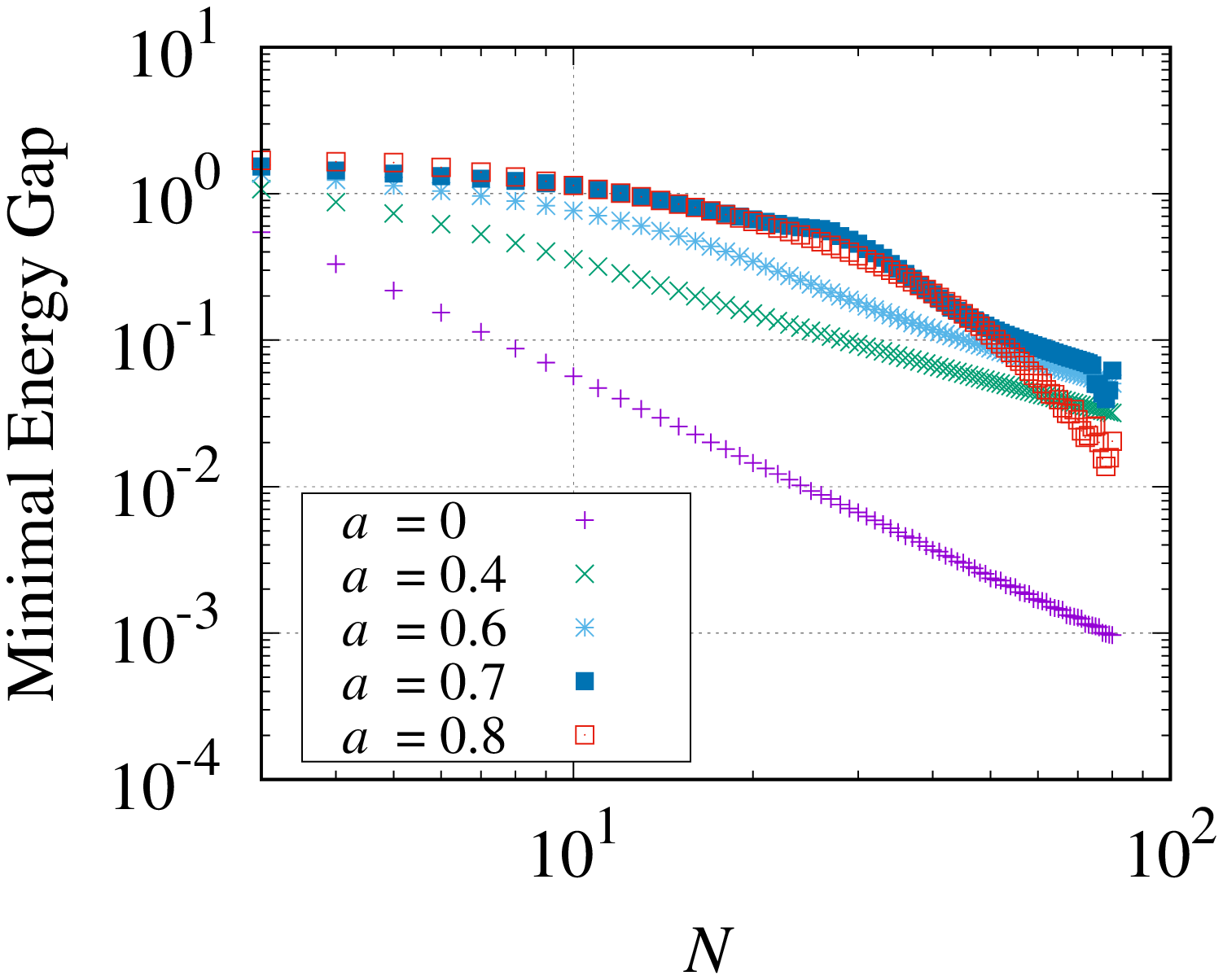}\label{fig:energy_gap2-b}}
    \subfigure[\ ]{\includegraphics[width=0.3\linewidth]{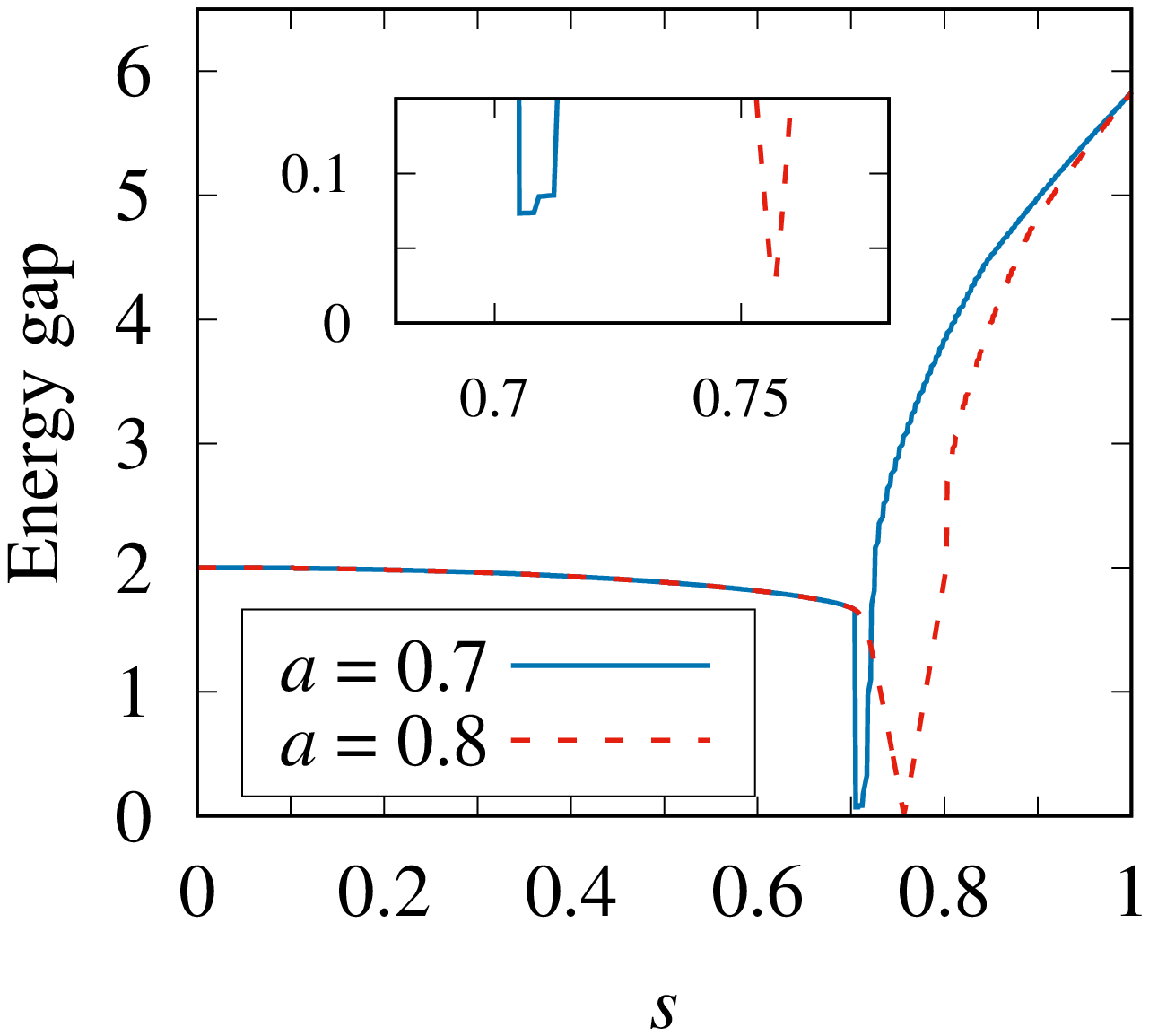}\label{fig:energy_gap2-c}}
  \caption{(a) The dashed lines show the schedule of $\tau$ expressed by Eq.~(\ref{eq:ramp}) in the phase diagram. The black sold line represents first-order phase transitions. (b) The minimal value of the energy gap against $N$ in a log-log scale as calculated by numerical diagonalization. (c) The energy gap for $N=70$ in two cases $a=0.7$ and $0.8$ of Eq.~(\ref{eq:ramp}). The inset shows the behavior around the phase transition point. All results shown are for $p=3$.}
   \label{fig:energy_gap2}
\end{figure*}

It is interesting and important to check the behavior of the minimum energy gap as a function of the system size.  As seen in Figs.~\ref{fig:energy_gap-a} and \ref{fig:energy_gap-c}, the minimum of the energy gap exists near the origin $\tau=s=0$ when there is no transition along the annealing path ($\tau=s$), whereas the minimum is at the critical point when such a transition exists along the path [Fig.~\ref{fig:energy_gap-b}].  We have chosen a series of paths as drawn in Fig.~\ref{fig:energy_gap2-a} to see the combined effects of the conventional path ($\tau=0$) and the inhomogeneous driving protocol ($\tau >0$). More explicitly, $\tau$ follows the schedule
\begin{align}
\label{eq:ramp}
\tau =
\left\{
\begin{array}{ll}
 0&\text{if}\ s<a, \\
 (s-a)/(1-a)&\text{if}\ s\geq a ,
\end{array}
\right.
\end{align}
with a control parameter $a$.
The path $\tau=s$ is reproduced with $a=0$, and the path with $a=0.4$ just touches the critical point at the end of the first-order line for $p=3$. For $a=0.8$, the path goes across the first-order transition point in the conventional homogeneous way $(\tau=0)$ and, only after the transition is crossed, the inhomogeneity sets in.  The minimal energy gap as a function of the system size, as shown in Fig.~\ref{fig:energy_gap2-b}, is seen to decrease polynomially for $a=0$ and $0.4$. The case of $a=0.8$ has an exponential decrease as expected from the existence of a first-order transition.  The remaining $a=0.6$ and 0.7 are marginal; a clear signal of an exponential decrease would show up only for larger system sizes than we studied here, $N=70$. In other words, the energy gap stays relatively large until the system size becomes very large if we choose a path along the $\tau=0$ axis (the conventional protocol) until just before a first-order phase transition is hit and then introduce the inhomogeneity. Figure~\ref{fig:energy_gap2-c} shows the $s$ dependence of the gap for $N=70$, the largest system size we studied.

\section{Nonideal cases}
\label{sec:practical_situation}

\begin{figure*}
  \centering
    \subfigure[\ ]{\includegraphics[width=0.3\linewidth]{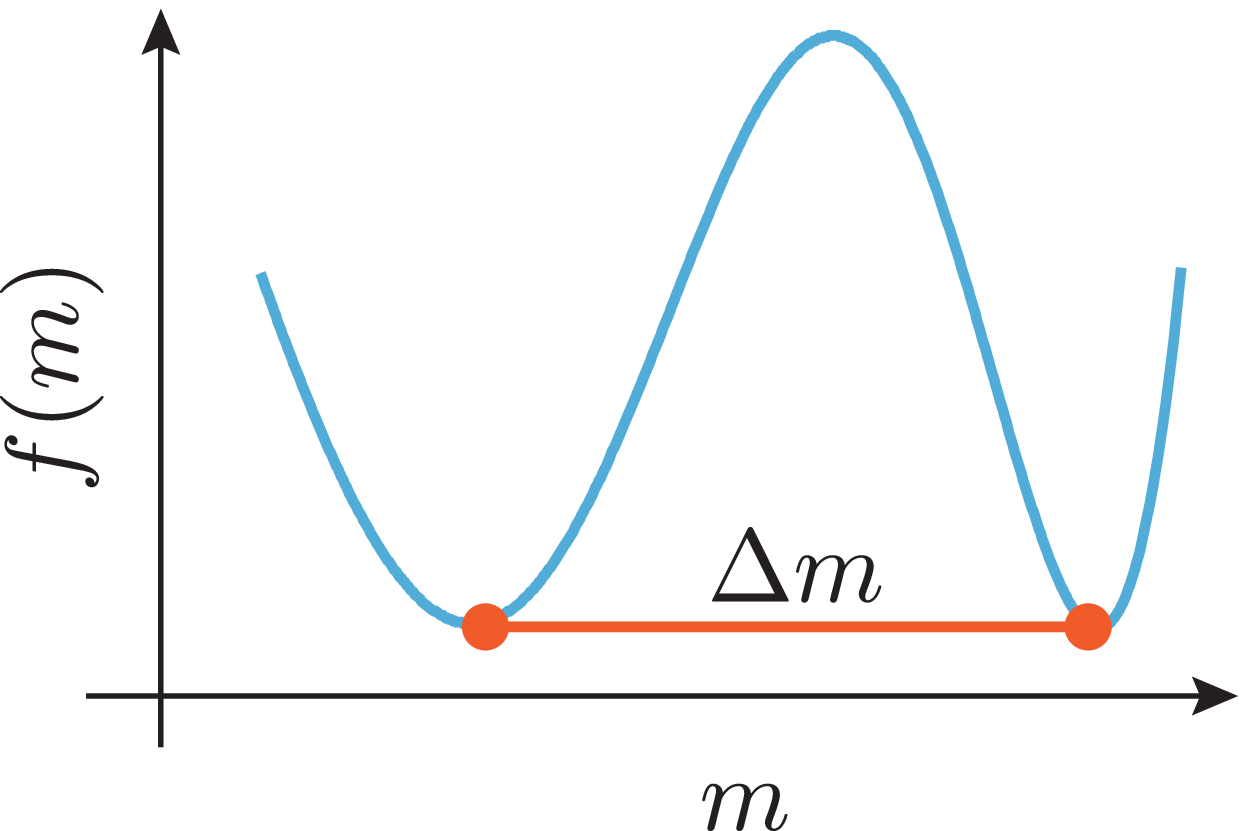}  \label{fig:order_parameter_illust}}
    \subfigure[\ ]{\includegraphics[width=0.34\linewidth]{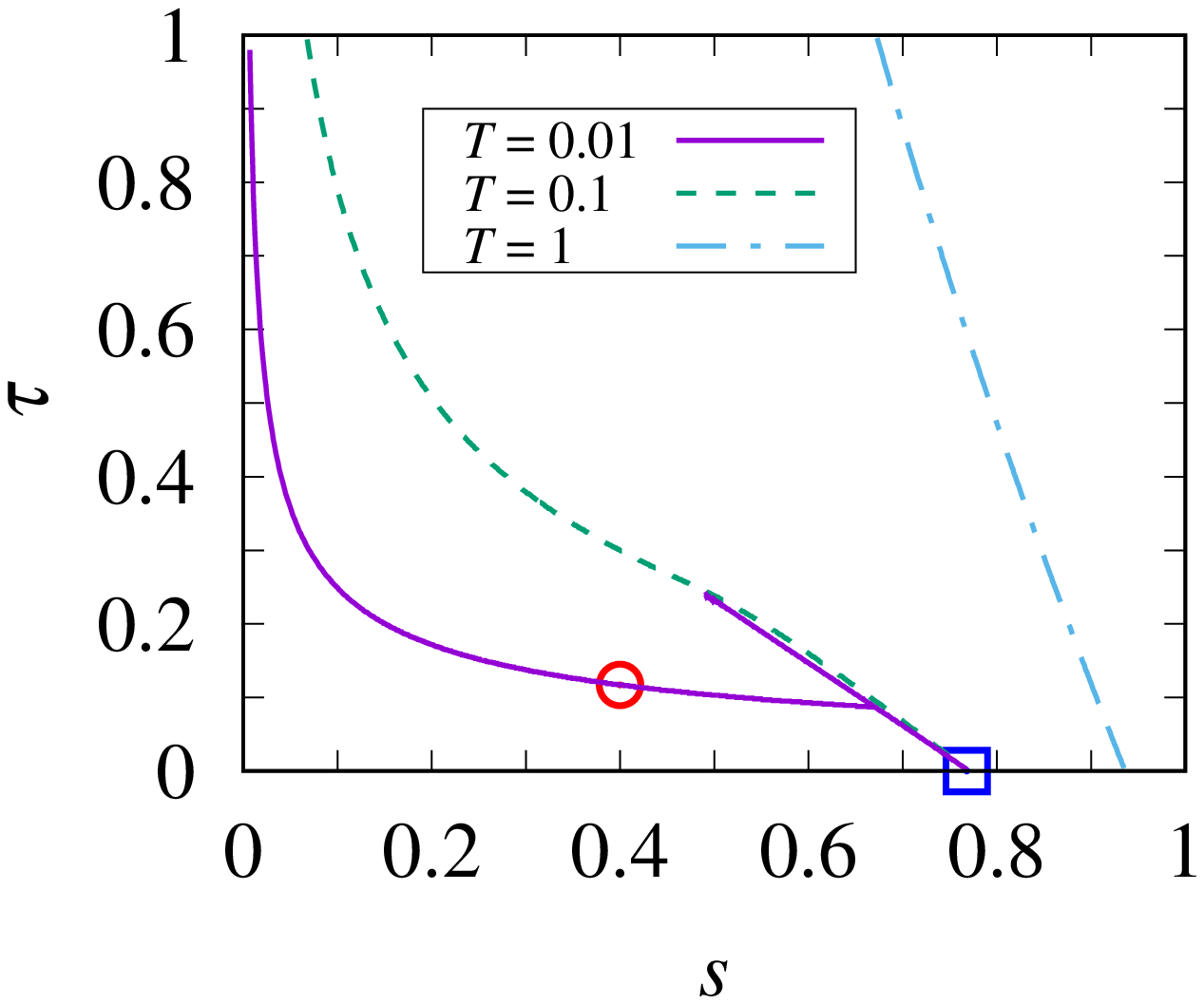}\label{fig:phase_diagram_finite_temp}}
    \subfigure[\ ]{\includegraphics[width=0.34\linewidth]{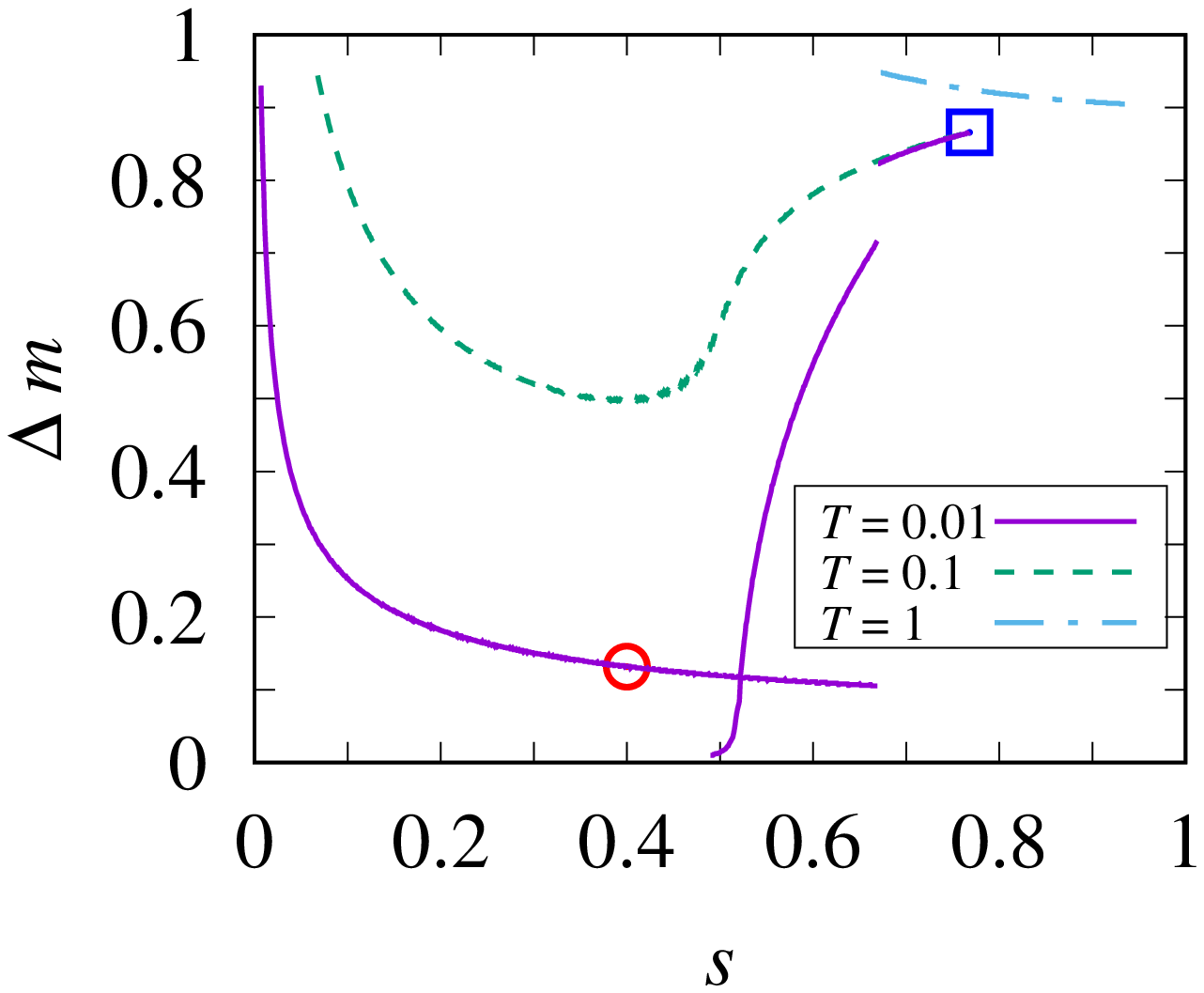}\label{fig:order_parameter_finite_temp}} 
    \subfigure[\ ]{\includegraphics[width=0.3\linewidth]{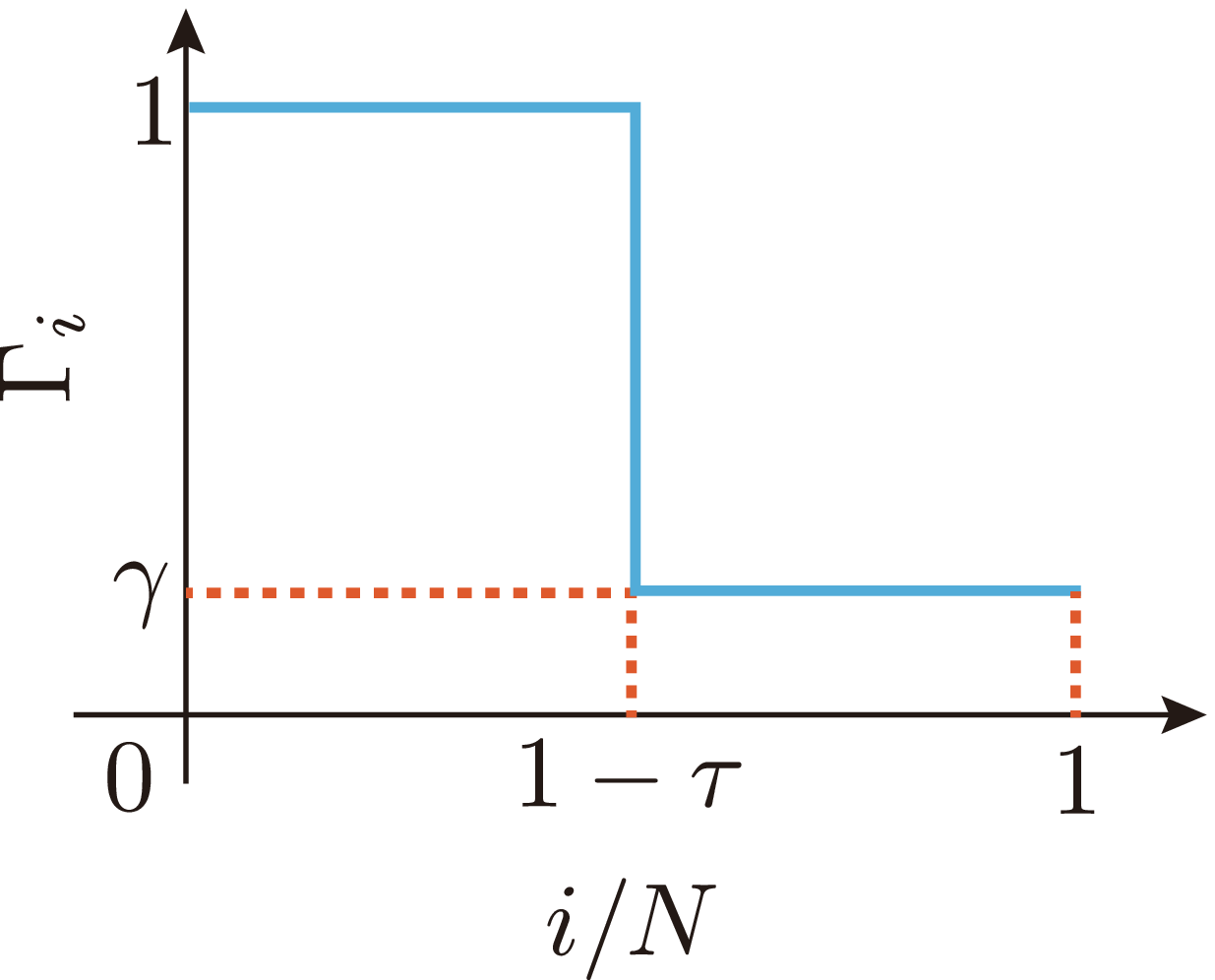}\label{fig:nonideal-a}}
    \subfigure[\ ]{\includegraphics[width=0.34\linewidth]{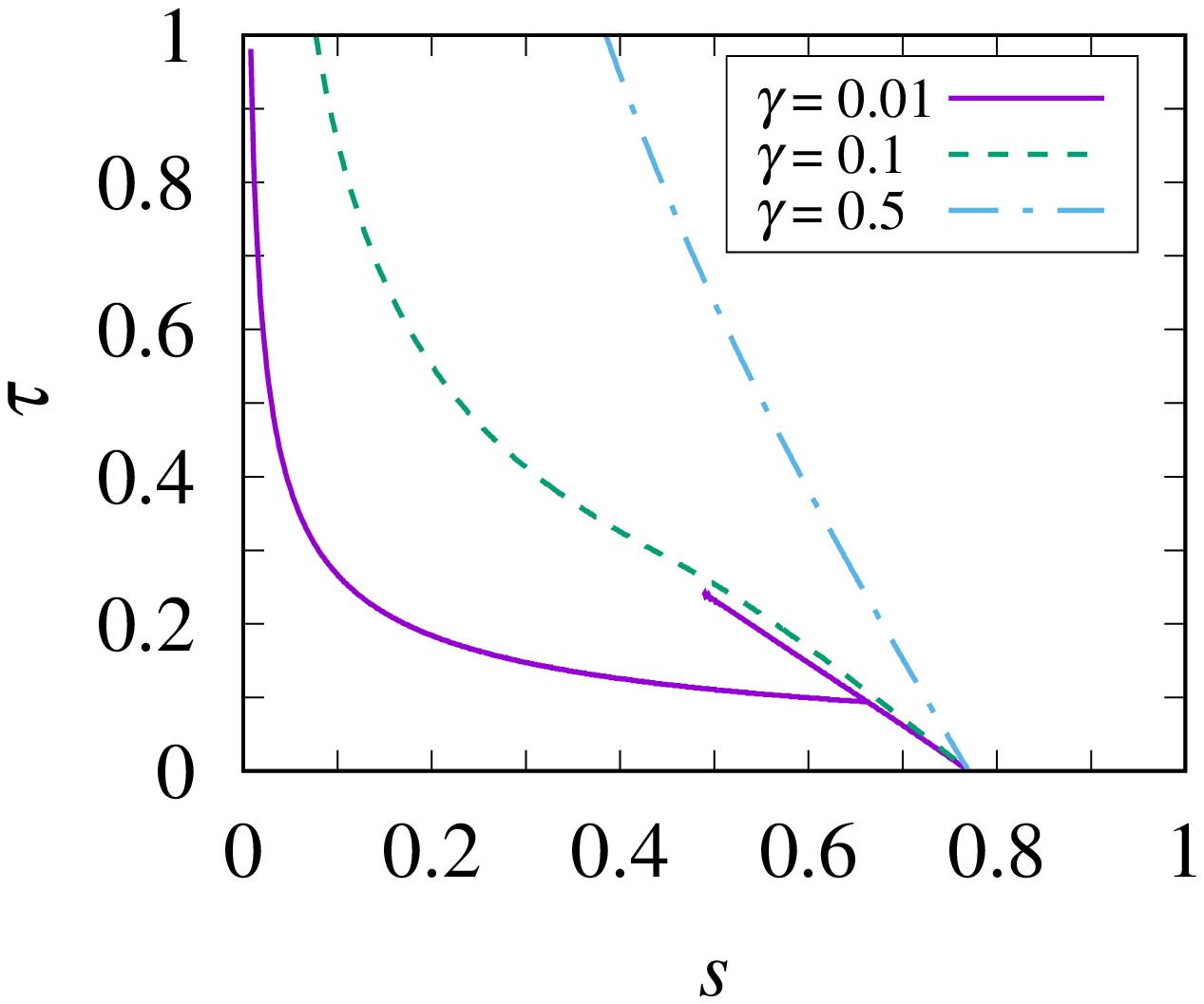}\label{fig:nonideal-b}}
    \subfigure[\ ]{\includegraphics[width=0.34\linewidth]{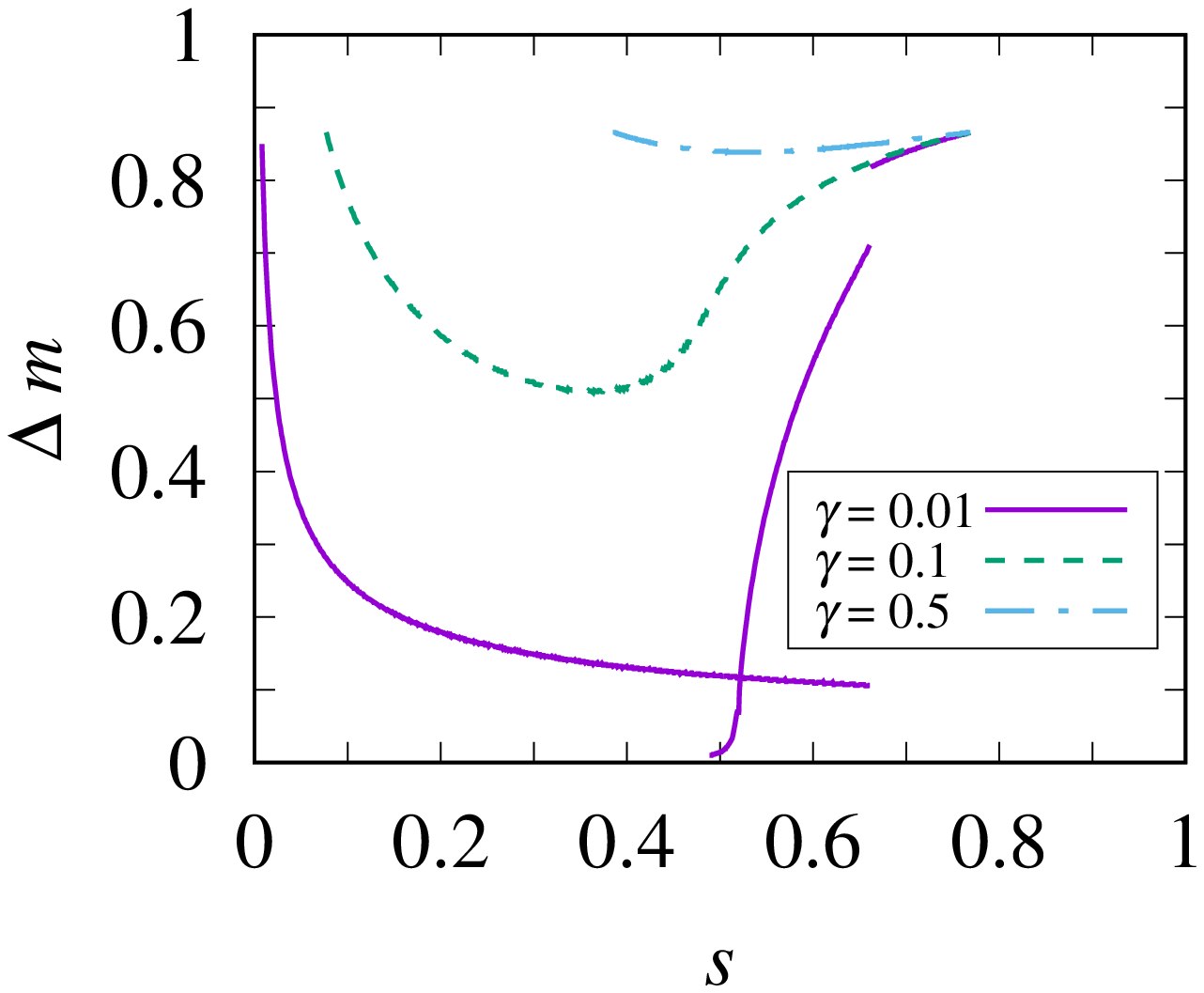}\label{fig:nonideal-c}}
    \caption{Two nonideal cases: (a)--(c) finite temperature and (d)--(f) incomplete turn-off. (a) Illustrative behavior of the free energy $f(m)$ and the jump $\Delta m$ in the order parameter at a first-order phase transition.
    (b) Finite-temperature phase diagram for $p=3$. All curves represent first-order phase transitions. The red circle and blue square correspond to the respective points in panel (c).
    (c) Jump in magnetization along the line of first-order transitions depicted in panel (b). 
    Symbols in red circle and blue square represent the respective points in the phase diagram of panel (b).
    (d) Amplitude of the transverse field $\Gamma_i$ of Eq.~(\ref{eq:unideal_schedule}). (e) Phase diagram for $p=3$. The curves represent first-order phase transitions. (f) Jump in magnetization $\Delta m$ along the first-order transition line. We note that (e) and (f) are remarkably similar to (b) and (c), though we do not presently have an explanation for this fact.
}
  \label{fig:nonideal}
\end{figure*}

The problem we studied in the previous section concerns the ideal case of zero temperature and a complete turning off of the transverse field at each site. In this section we relax some of these restrictions in order to see what happens under nonideal circumstances.

\subsection{Phase transition at finite temperature}

It is straightforward to draw the phase diagram at finite (but low) temperature from the free energy and the self-consistent equation, Eqs.~(\ref{eq:free_energy_finite_temp}) and (\ref{eq:SCE1}).  The result is depicted in Fig.~\ref{fig:phase_diagram_finite_temp} with the annealing schedule of Eq.~(\ref{eq:inhomogeneous_schedule2}) kept intact.

As seen in the case of $T=0.01$, a new line of first-order transitions appear at low but finite temperature in addition to the line that already exists at $T=0$.  This new line of first-order transitions merges with the existing line at $T=0$ as the temperature rises, as observed in the cases of $T=0.1$ and $1$.

To understand what happens at this new transition line, it is useful to fix $\tau$ at a low but finite value and consider the system behavior as $s$ is increased.  For small $s$, the influence of the ferromagnetic interactions in the cost function $\hat{H}_0$ is weak and the system is disordered (magnetization $m=0$) due to thermal fluctuations at finite temperature.  As $s$ increases, the system is driven into the ferromagnetic phase $(m>0)$, which is heralded by the new first-order transition appearing in the finite-temperature phase diagram. For small $\tau$, the other first-order transition that already existed at $T=0$ causes a jump in magnetization from a small value to a larger value.  If we reduce the temperature from a small but finite value toward zero, the location of this first-order transition comes closer to the $s=0$ axis until it merges with the $s=0$ axis in the zero-temperature limit. In other words, at $T=0$, the system becomes ordered ($m>0$) as soon as a finite value of $s$ is introduced, as long as $\tau>0$. 

The structure of the phase diagram makes it impossible to avoid a first-order transition at finite temperature when one starts from the origin $s=\tau=0$ and proceeds toward the goal at $s=\tau=1$.  Nevertheless, the inhomogeneous driving protocol leads to quantitative improvements, if not qualitative, over its homogeneous counterpart.  To see this, we calculate the jump in magnetization $\Delta m$ along the line of first-order transitions. The jump represents the width of a free energy barrier at a first-order transition as illustrated in Fig.~\ref{fig:order_parameter_illust}. Thus, a decrease of the jump $\Delta m$ enhances the quantum tunneling rate through the free energy barrier quantitatively though the exponential dependence of the tunneling rate on the system size is unchanged.\footnote{The connection between the free-energy barrier width and tunneling rates can be made quantitative using the instanton method; see, e.g., \cite{jorg2010energy,Matsuura:2016aa}.}

Figure \ref{fig:order_parameter_finite_temp} shows the result. The red circle denotes the value of the jump at the point marked by the same red circle in the phase diagram of Fig.~\ref{fig:phase_diagram_finite_temp}, as a representative example of the system behavior under inhomogeneous field. The same is true for the blue square in Figs.~\ref{fig:order_parameter_finite_temp} and \ref{fig:phase_diagram_finite_temp}, this being for the conventional homogeneous annealing case. In general, any point on the purple curve $T=0.01$ in Fig.~\ref{fig:order_parameter_finite_temp} shows $\Delta m$ at the corresponding first-order transition point on the purple curve ($T=0.01$) in Fig.~\ref{fig:phase_diagram_finite_temp}. It is clearly seen that the jump is reduced at $T=0.01$ and 0.1  for most values of $s$ in comparison with the homogeneous case marked by the blue square. 
We may therefore conclude that inhomogeneous driving is advantageous to standard homogenous driving in that it enhances the tunneling rate even when a first-order transition is unavoidable, as in the present nonideal (finite temperature) situation.

\subsection{Different types of inhomogeneity}

Let us next consider the case with a nonvanishing final value of the transverse field, at $T=0$.  We expect this prescription to induce a similar behavior to the finite-temperature case as the nonvanishing transverse field may disorder the system after the field is turned off incompletely.

The formal definition of the transverse field is now
\begin{align}
\label{eq:unideal_schedule}
\Gamma_i=
\begin{cases}
1 &\text{for}\ \  0\leq i/N\leq 1-\tau, \\
\gamma &\text{for}\ \  1-\tau<i/N\leq1,
\end{cases}
\end{align}
where a small transverse field ($0<\gamma<1$) remains after an incomplete turn-off [Fig.~~\ref{fig:nonideal-a}]. It is easy to show from Eq.~(\ref{eq:free_energy_zero_temp}) that the free energy at zero temperature becomes 
\begin{align}
f(m)=&s(p-1) m^p -(1-\tau) \sqrt{(spm^{p-1})^2+1}\notag \\
&-\tau \sqrt{(spm^{p-1})^2+\gamma^2},
\end{align}
which is to be compared with Eq.~(\ref{eq:free_energy}). The phase diagram and the behavior the order parameter can be derived from this free energy.

Figure~\ref{fig:nonideal-b} is the phase diagram and Fig.~\ref{fig:nonideal-c} is the jump in magnetization $\Delta m$ along the transition line. The qualitative similarity to the finite temperature case depicted in Figs.~\ref{fig:phase_diagram_finite_temp} and \ref{fig:order_parameter_finite_temp} is striking. We conclude that quantum fluctuations induced by a small but finite $\gamma$ indeed play a similar role as the temperature effects.

\begin{figure}
  \centering
\subfigure[\ ]{\includegraphics[width=0.7\linewidth]{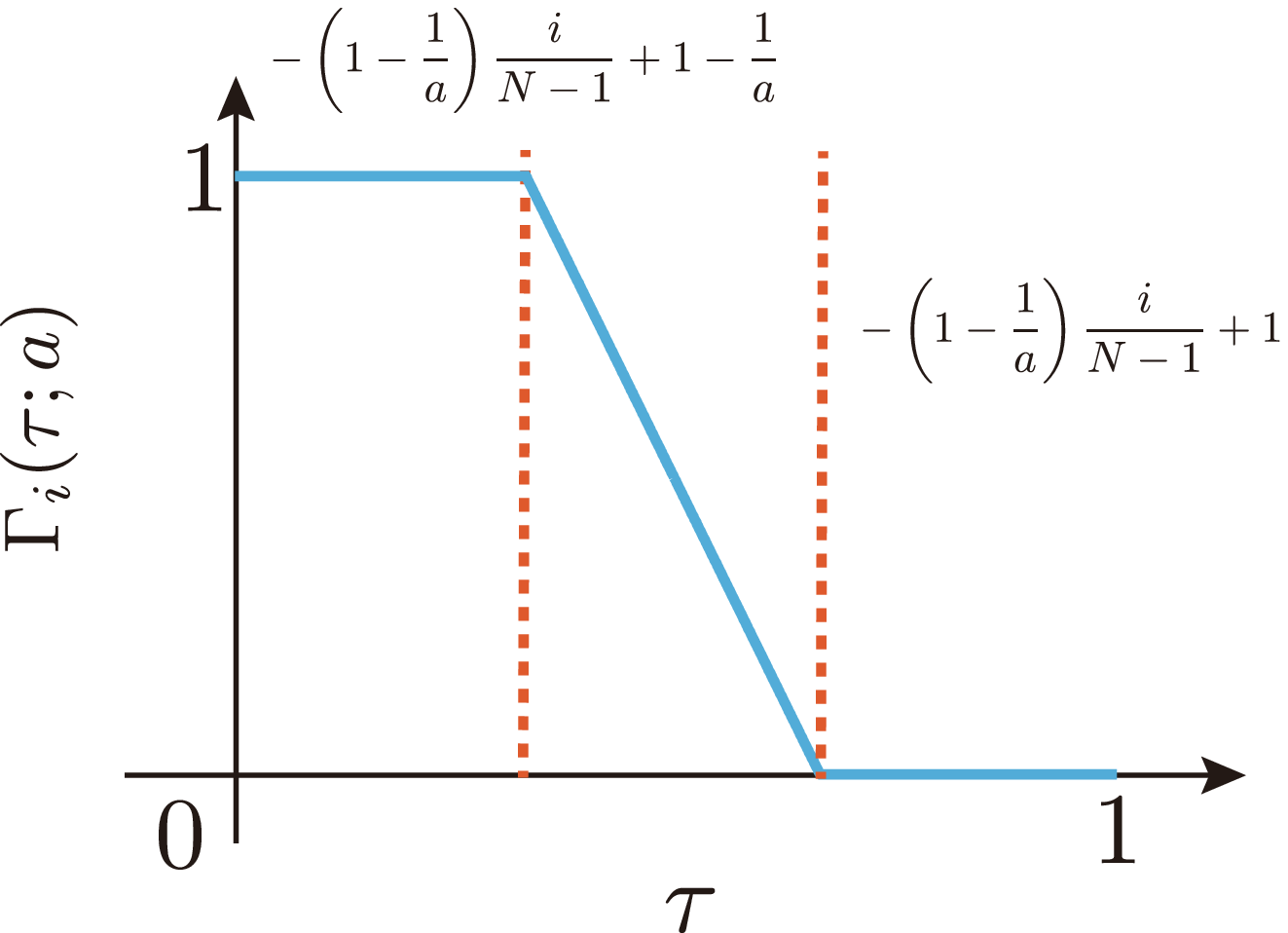}\label{fig:finiteslope-a}}\\
\subfigure[\ ]{\includegraphics[width=0.8\linewidth]{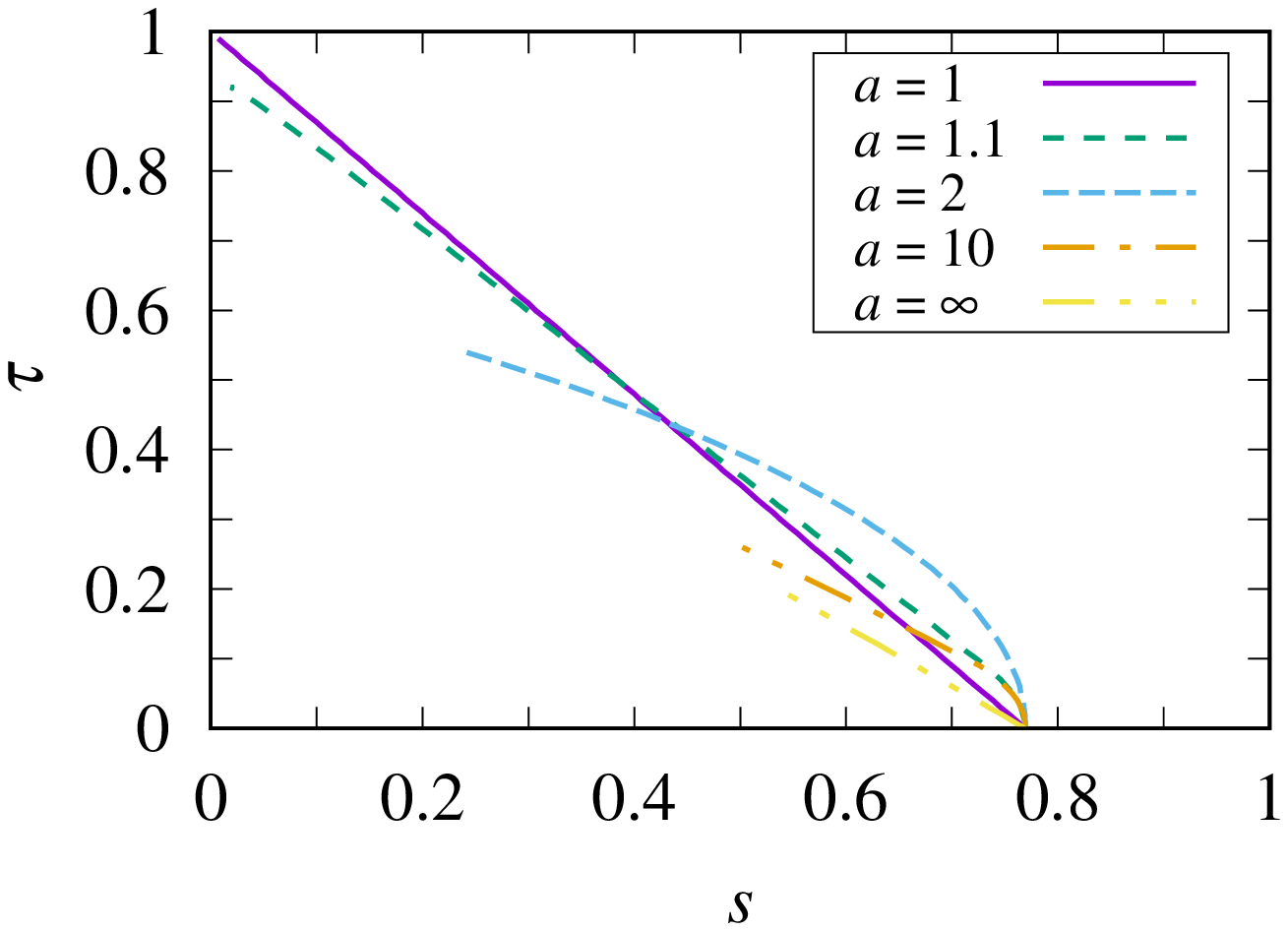}\label{fig:finiteslope-b}}
  \caption{(a) The field amplitude $\Gamma_i(\tau ; a)$ of Eq.~(\ref{eq:finite_solope_field}). (b) Phase diagram for $p=3$ for several values of $a$.}
  \label{fig:finiteslope}
\end{figure}

As the second example, we study the following function \cite{rams2016inhomogeneous},
\begin{widetext}
\begin{align}
&\Gamma_i \left(\tau ; a\right) =
  \begin{cases}
  0 &\text{for}\ \tau > -\left(1-\frac{1}{a}\right)\frac{i}{N-1} + 1\\
  a\left(1- \tau\right)-(a-1)\frac{i}{N-1} & \text{otherwise}\\
  1 &\text{for}\ \tau < -\left(1-\frac{1}{a}\right)\frac{i}{N-1} + 1 - \frac{1}{a}
  \end{cases},
  \label{eq:finite_solope_field}
\end{align}
\end{widetext}
which is drawn in Fig.~\ref{fig:finiteslope-a}. The parameter $a$ controls the slope that interpolates two values $\Gamma_i=0$ and 1. The limit $a\to 1$ corresponds to the homogeneous field, whereas $a\to\infty$ is the simple step function of Eq.~(\ref{eq:inhomogeneous_schedule2}). 

The zero-temperature free energy is derived from Eqs.~(\ref{eq:free_energy_zero_temp}) and (\ref{eq:finite_solope_field}) and reads
\begin{align}
f(s,\tau;m) =& (p-1)sm^p + x_1\sqrt{(spm^{p-1})^2 + 1} \notag \\
&+ G(\Gamma_0) - G(\Gamma_1) + (1-x_0)spm^{p-1} ,
\end{align}
where
\begin{align}
x_1 &= \begin{cases}
    1-\frac{a}{a-1} &\text{for}~ \tau < 1-\frac{1}{a}\\
    0 &\text{for}~ 1-\frac{1}{a} \leq \tau
    \end{cases},\\
x_0 &= \begin{cases}
    1 &\text{for}~\tau < \frac{1}{a}\\
    \frac{a}{a-1}(1-\tau) &\text{for}~\frac{1}{a} \leq \tau
    \end{cases},\\
\Gamma_1 &= \begin{cases}
    1 &\text{for}~\tau < 1-\frac{1}{a}\\
    a(1-\tau) &\text{for} ~1-\frac{1}{a} \leq \tau
    \end{cases}, \\
\Gamma_0 &= \begin{cases}
    1-a\tau &\text{for}~\tau < \frac{1}{a}\\
    0 &\text{for}~\frac{1}{a} \leq \tau
    \end{cases},
\end{align}
and
\begin{align}
G(\Gamma) =& - \frac{1}{2(a-1)}\left\{ \Gamma \sqrt{(spm^{p-1})^2 + \Gamma^2} \right. \notag \\
&\left. + (spm^{p-1})^2 \ln \left(\sqrt{(spm^{p-1})^2+\Gamma^2}+\Gamma\right)
\right\}.
\end{align}
Figure~\ref{fig:finiteslope-b} is the resulting phase diagram. It can be seen that paths exist that avoid first-order transitions when the inhomogeneity is turned on, i.e., $a>1$. 

As mentioned earlier, Ref.~\cite{rams2016inhomogeneous} discusses inhomogeneous annealing for a weakly disordered ferromagnetic one-dimensional chain. It is not straightforward to compare our results with theirs, since this is a very different problem with its own characteristics such as a low cost of domain formation. Nevertheless, the conclusion common to both this work and Ref.~\cite{rams2016inhomogeneous} is that inhomogeneous driving is useful for reaching better solutions.

\subsection{Longitudinal random field}

We next consider the case with random longitudinal fields: 
\begin{align}
\hat{H}_0=-N\left(\frac{1}{N}\sum_{i=1}^{N}\hat{\sigma}_i^z\right)^p-\sum_{i=1}^{N} h_i \hat{\sigma}_i^z,
\end{align}
where each $h_i$ is drawn from the bimodal or the Gaussian distribution:
\begin{subequations}
\begin{align}
\label{eq:binary_random}
P_b(h_i) &= \frac{1}{2}\left[\delta(h_i+h_0)+\delta(h_i-h_0)\right], \\
\label{eq:gauss_random}
P_g(h_i) &= \frac{1}{\sqrt{2\pi \sigma^2}} e^{-h_i/2\sigma^2}.
\end{align}
\end{subequations}
It is noteworthy that the introduction of nonstoquasticity into the Hamiltonian of the $p$-spin model without random field removes a first-order phase transition for $p>3$ \cite{seki2012quantum,seoane2012many,nishimori2017exponential} whereas the same idea fails if random longitudinal field exists  \cite{ichikawa_thesis}. Thus, this model with random field is a test bed to compare the performance of inhomogeneous driving and that of the nonstoquastic Hamiltonian. 

The computation of the free energy proceeds as before, and the result for $T=0$ is
\begin{align}
f(m)=&s(p-1) m^p -\left[ \int_0^1 dx  \sqrt{(spm^{p-1}+h)^2+\Gamma(x)^2}\right],
\end{align}
where the brackets $[\cdots]$ denote the average over the distribution of the random field variable denoted as $h$, and we have used the law of large numbers,
\begin{align}
    \lim_{N\to\infty}\frac{1}{N}\sum_{i=1}^N ( \cdots )=[(\cdots)].
\end{align}
\begin{figure}
  \centering
\subfigure[\ ]{\includegraphics[width=0.8\linewidth]{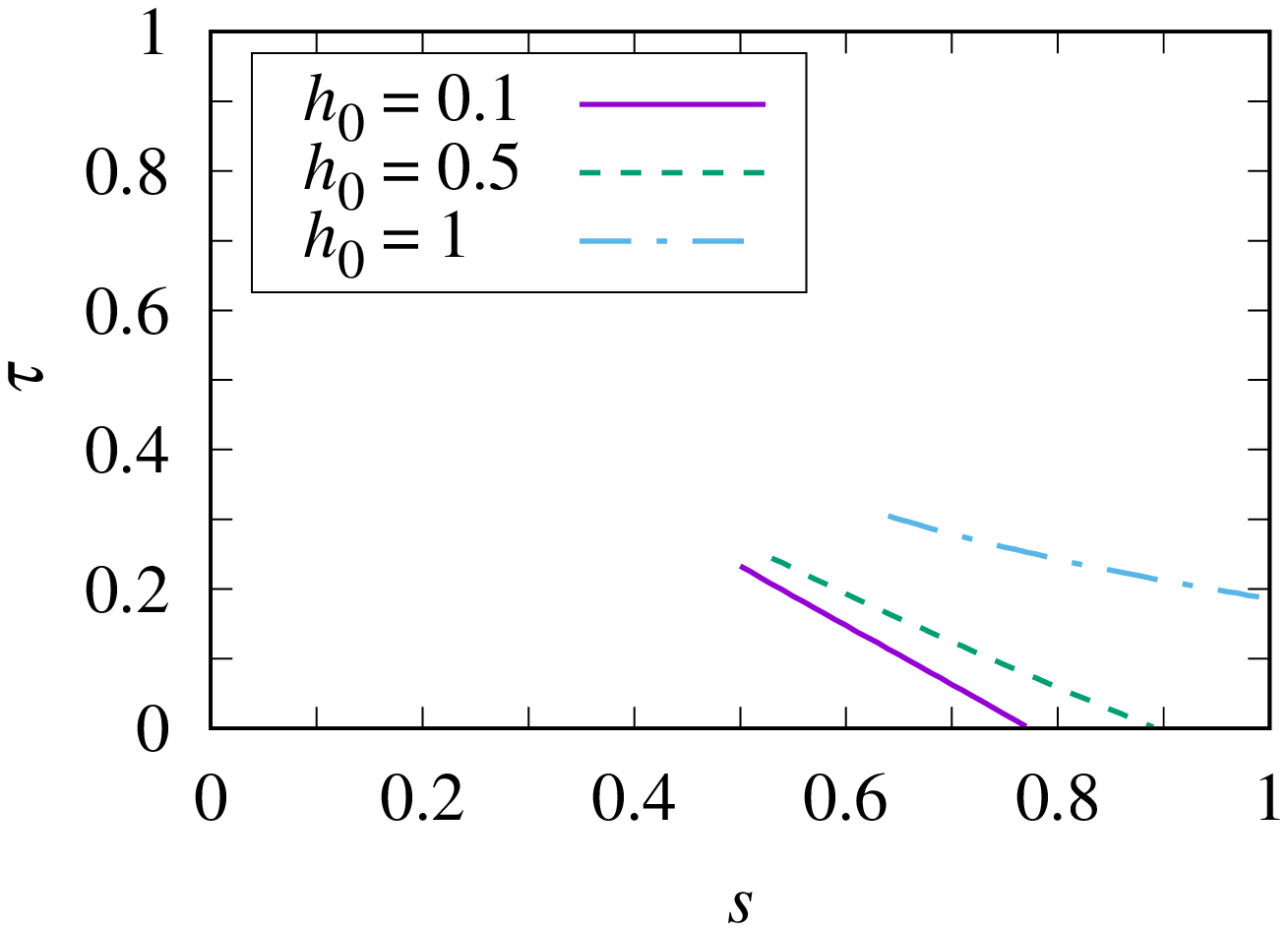}\label{fig:phase_diagram_random-a}}\\
\subfigure[\ ]{\includegraphics[width=0.8\linewidth]{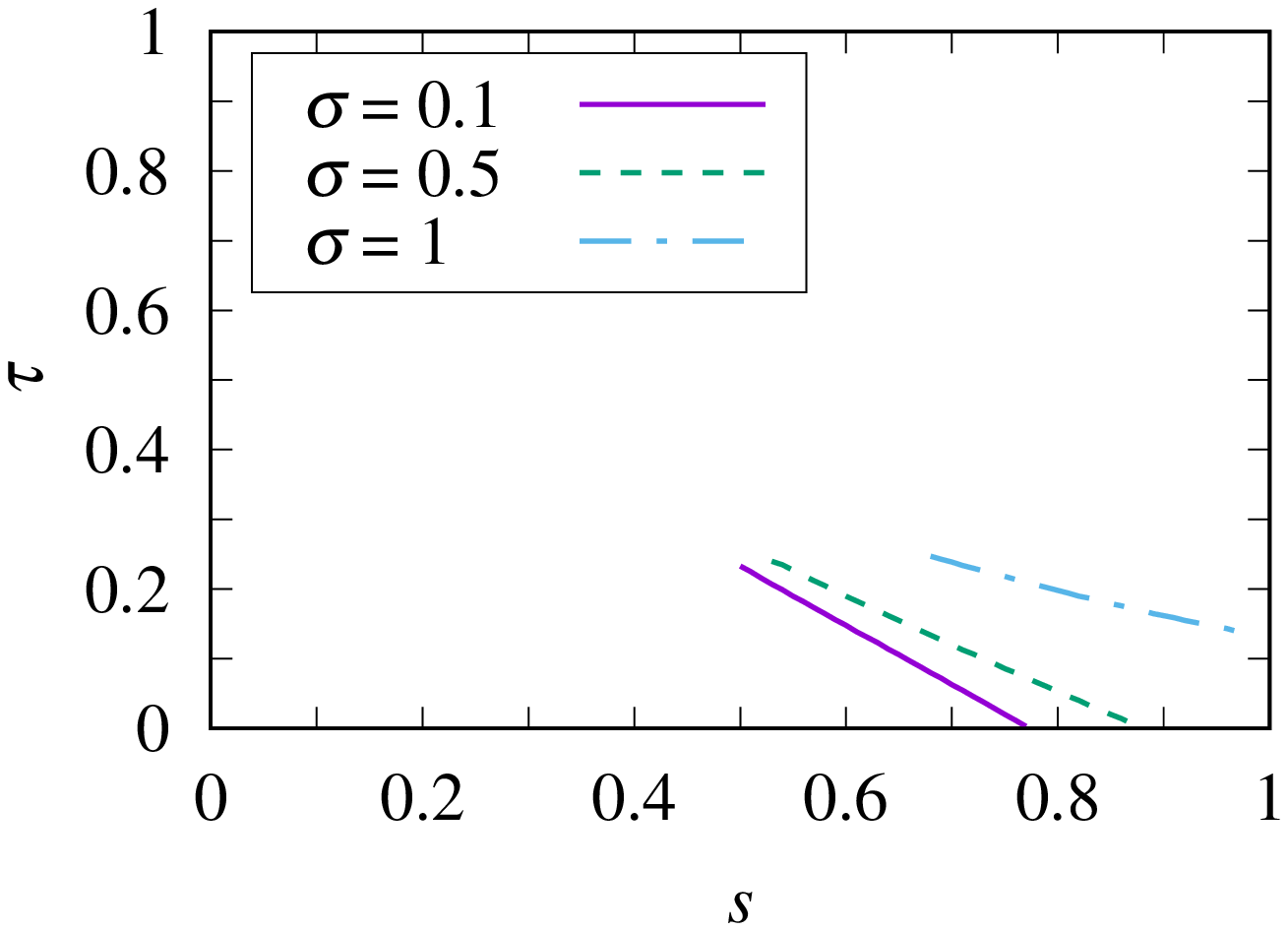}\label{fig:phase_diagram_random-b}}
  \caption{(a) Phase diagram for bimodal random longitudinal fields with strengths $h_0=0.1,\ 0.5$, and $1$. (b) Phase diagram for Gaussian random longitudinal fields with standard deviations $\sigma=0.1,\ 0.5$, and $1$. All lines are for first-order phase transitions. All the data are for $p=3$. 
}
  \label{fig:phase_diagram_random}
  \end{figure}
Figure~\ref{fig:phase_diagram_random} shows the phase diagram for the simple inhomogeneity of Eq.~(\ref{eq:inhomogeneous_schedule}). Panels (a) and (b) are for the bimodal and Gaussian distributions, respectively. In both cases we see that the inhomogeneous transverse field eliminates first-order phase transitions.  This leads to the interesting conclusion that the present method of inhomogeneous driving of the transverse field is more powerful for the removal of first-order transitions than the introduction of non-stoquastic Hamiltonians, at least for the $p$-spin model under random longitudinal fields.

\section{Comparison with classical models}
\label{sec:classical}
It is useful to compare the results of the previous sections with those of the classical counterparts of QA.  Here we focus on two classical models: simulated annealing \cite{Kirkpatrick1983} and spin vector Monte Carlo (SVMC) \cite{Shin2014}.

\subsection{Simulated Annealing with an Inhomogeneous Temperature Schedule}

A ``limited quantum speedup" is a speedup of quantum annealing relative to its classical counterparts, such as simulated annealing \cite{ronnow2014defining}.  Indeed, it was through this viewpoint that the concept of quantum annealing was proposed in Ref.~\cite{kadowaki1998quantum}. We therefore study the classical Ising model with an inhomogeneous driving parameter, i.e., the (inverse) temperature in simulated annealing. We consider the $p$-spin model under random local fields:
\begin{align}
H=-N \left(\frac{1}{N}\sum_{i=1}^{N}\beta_i \sigma_i\right)^p-\sum_{i=1}^N \beta_i h_i \sigma_i,
\end{align}
where $\sigma_i(=\pm 1)$ is a simple classical Ising variable and $\beta_i$ is the inhomogeneous (site-dependent) inverse temperature. It is to be noted that we take the above Hamiltonian to be dimensionless, corresponding to the product $\beta H$, where $\beta$ is the (homogeneous) inverse temperature. The site-dependent temperature $T_i=1/\beta_i$ is also dimensionless. The random field $h_i$ follows the bimodal or the Gaussian distribution.

The partition function can be calculated as
\begin{align}
Z =& \Tr e^{-H} \notag\\
=& \Tr \,\int dm\ \delta \left(Nm-\sum_{i=1}^N \beta_i \sigma_i\right)e^{Nm^p+\sum_{i=1}^N\beta_i h_i\sigma_i} \notag\\
=& \Tr \int dm\, d\tilde{m}\, e^{-\tilde{m}\left(Nm-\sum_{i=1}^N \beta_i \sigma_i \right)+Nm^p+\sum_{i=1}^N\beta_i h_i\sigma_i} \notag\\
=& \int dm\, d\tilde{m}\, e^{-Nm\tilde{m}+Nm^p+\sum_{i=1}^N\ln 2\cosh \beta ( \tilde{m}+h_i)} .
\end{align}
The saddle-point condition with respect to $m$ is $\tilde{m}=pm^{p-1}$. Then the free energy per site is
\begin{align}
f=(p-1)m^p-\frac{1}{N}\sum_{i=1}^N \ln 2 \cosh \beta_i (pm^{p-1}+h_i).
\end{align}
Under the inhomogeneous protocol we decrease the local temperature or increase the inverse temperature $\beta_i$ sitewise. Suppose that $\beta_i=0$ for $i = 1,2,\cdots ,N(1-\tau)$ and $\beta_i = \beta_0$ for $i = N(1-\tau)+1,\cdots ,N$.  In other words, the local temperature $T_i=1/\beta_i$ has been decreased from $\infty$ to $1/\beta_0$ for $N\tau$ spins and is kept $\infty$ for the remaining $N(1-\tau)$ spins. Thus, as we increase $\tau$ from $0$ to $1$, the fraction of sites with low temperature increases. Under this prescription, the free energy per spin becomes
\begin{align}
f=(p-1)m^p-\tau \left[\ln 2 \cosh \beta_0 (pm^{p-1}+h_i)\right] + \text{const}.
\end{align}

\begin{figure}
  \centering
  \includegraphics[width=0.49\linewidth]{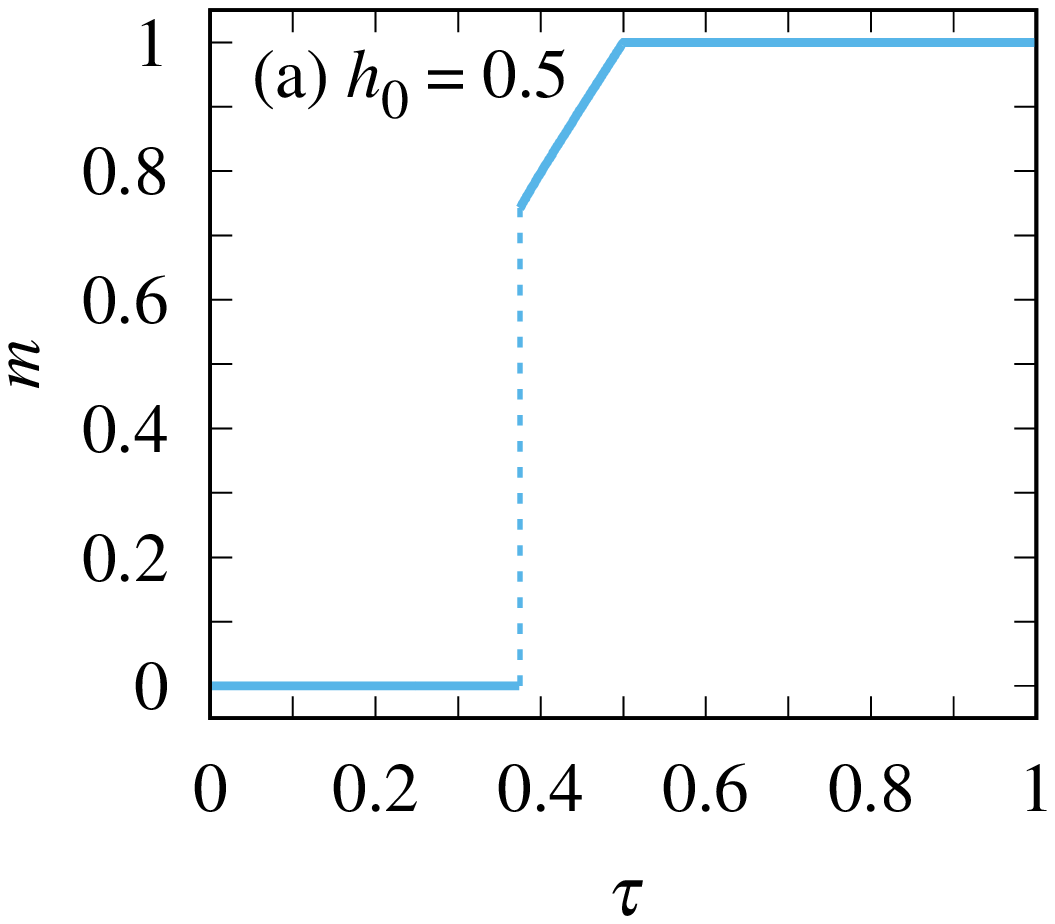}
  \includegraphics[width=0.49\linewidth]{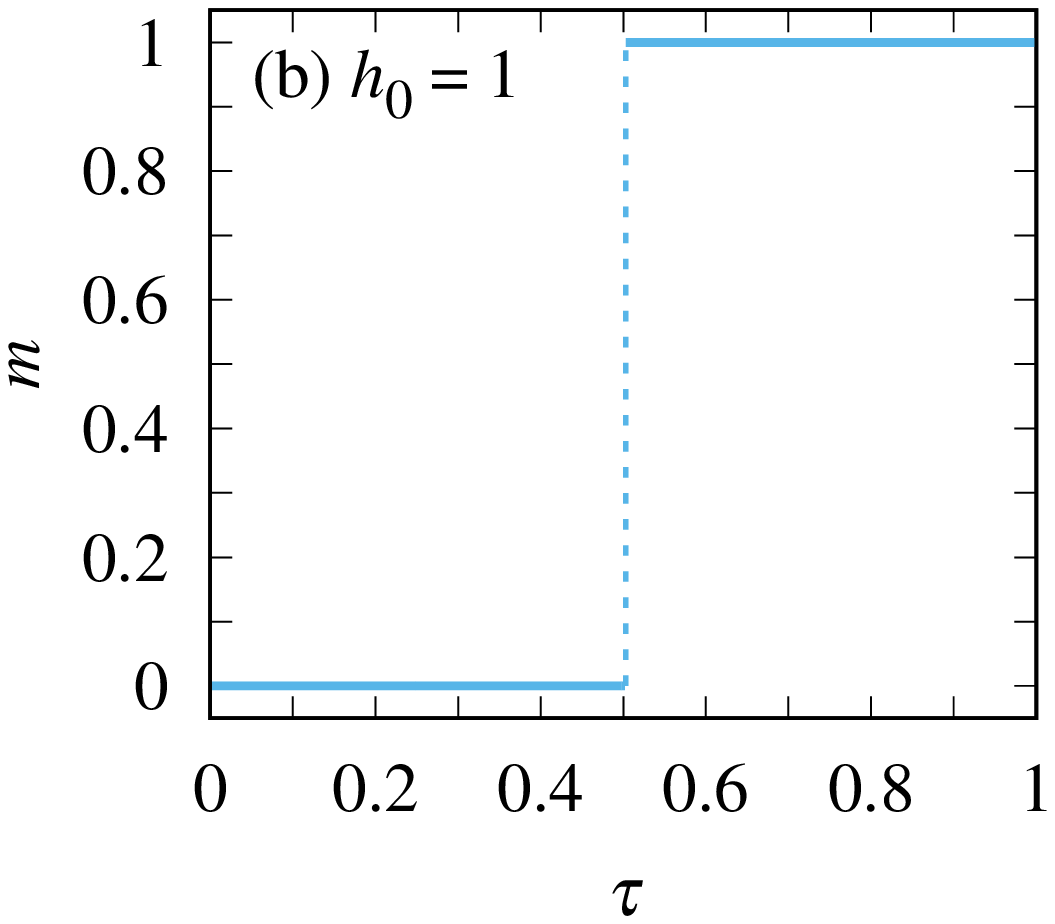}
  \caption{Behavior of the order parameter in simulated annealing with inhomogeneous temperature driving. We choose $p=3$ and $\beta_0=2$ and the bimodal distribution of random local fields.}
  \label{fig:inhomogeneous_SA}
\end{figure}
Figure \ref{fig:inhomogeneous_SA} shows the order parameter $m=(1/N)\sum_{i=1}^N \beta_i \sigma_i$ evaluated from the free energy. Here the amplitude of the bimodal distribution of random fields is chosen as (a) $h_0=0.5$ and (b)  $h_0=1$. This figure shows that the first-order phase transition does not disappear in simulated annealing under inhomogeneous temperature driving, since the order parameter has a discontinuity. We found essentially the same behavior for any combination of the parameters, $p(\ge 3),\  \beta_0,\ h_0$, and $\tau$, as long as $\beta_0$ or $h_0$ is not too large, in which cases the final state belongs to the same paramagnetic phase as the initial one, and therefore no phase transition can ever happen.  The same holds for the Gaussian distribution of random fields.  We therefore conclude that inhomogeneous temperature driving of simulated annealing is incapable of removing a first-order transition (at least for the $p$-spin model), in contrast to the corresponding quantum case.

\subsection{Spin vector Monte Carlo}

In this section we consider the spin vector Monte Carlo (SVMC) algorithm, in which one replaces $\hat{\sigma}_i^x$ and $\hat{\sigma}_i^z$ by $\sin \theta_i$ and $\cos \theta_i$, respectively, and applies Metropolis moves to update the angles. This algorithm was developed as a classical model for the D-Wave processors~\cite{Shin2014}, and has been the subject of scrutiny in this context \cite{Albash:2014if,q-sig2,PAL:14,Boixo:2014yu,Mishra:2018aa}. It can be derived as the semiclassical limit of the spin-coherent states path integral, so that it can be understood as a mean-field approximation of the simulated quantum annealing (SQA) algorithm~\cite{Albash:2014if,Crowley:2014qp}. We therefore anticipate that it will be a close approximation to our mean-field solution of the $p$-spin model as well.

\begin{figure}
  \centering
  \includegraphics[width=0.8\linewidth]{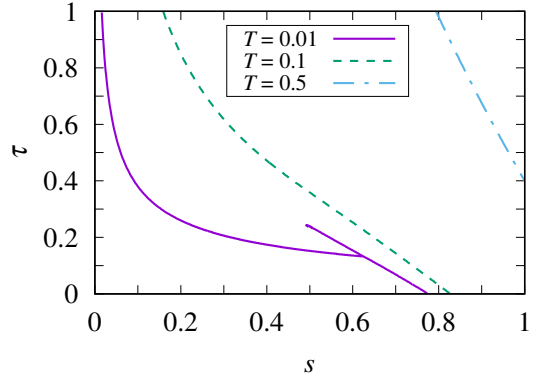}
  \caption{Phase diagram for $p=3$ for the SVMC model.}
  \label{fig:SVMC}
\end{figure}

In the context of the $p$-spin model with an inhomogeneous transverse field, the Hamiltonian is rewritten in the SVMC model as
\begin{align}
H(s) = -sN \left(\frac{1}{N} \sum_{i=1}^{N} \cos \theta_i\right)^p -\sum_{i=1}^{N} \Gamma_i \sin \theta_i.
\end{align}
The partition function is calculated as
\begin{align}
Z=& \Tr e ^{-\beta  H(s)} \notag \\
=& \Tr \int dm \,\delta \left(Nm - \sum_{i=1}^{N} \cos \theta_i\right) e^{\beta(sNm^p+\sum_{i=1}^N \Gamma_i \sin \theta_i)} \notag \\
=&  \Tr \int dm \int d\tilde{m}\, e^{i(Nm-\sum_{i=1}^{N} \cos \theta_i)\tilde{m}+\beta (sN m^p +\sum_{i=1}^{N}\Gamma_i \sin \theta_i) }.
\end{align}
The saddle-point condition for $m$ is $i\tilde{m}+\beta s p m^{p-1}=0$. The trace over the angles is straightforwardly evaluated as
\begin{align}
&\Tr \exp\left[-i\tilde{m} \sum_{i=1} ^{N}\cos \theta_i + \beta \sum_{i=1} ^{N} \Gamma_i \sin \theta_i\right] \notag \\
&= \prod_{i=1}^N  \int_0^{2\pi} d \theta_i \exp \left[\beta s p m^{p-1} \cos \theta_i + \beta \Gamma_i  \sin \theta_i \right]  \notag \\
&=  \prod_{i=1}^N  2 \pi I_0\left(\beta \sqrt{(spm^{p-1})^2+\Gamma_i^2} \right),
\end{align}
where $I_n(x)$ is the modified Bessel function of the first kind. Then we have:
\begin{align}
Z =&\int dm \exp\Bigg[-\beta (p-1) s N m^p \notag \\
&+\sum_{i=1}^N  \ln\left\{2\pi I_0 \left(\beta \sqrt{(spm^{p-1})^2+\Gamma_i^2} \right) \right\} \Bigg].
\end{align}
Thus the free energy per spin is
\begin{align}
\label{eq:free_energy_SVMC}
f 
=& s (p-1) m^p \notag \\
&-\frac{1}{\beta N}\sum_{i=1}^N \ln \left\{2\pi I_0\left( \beta \sqrt{(spm^{p-1})^2+\Gamma_i^2}\right) \right\} \notag \\
=& s (p-1) m^p \notag \\
&-\frac{1}{\beta}\int_{0}^1 dx \ln \left\{2\pi I_0\left( \beta \sqrt{(spm^{p-1})^2+\Gamma(x)^2}\right) \right\},
\end{align}
where we replaced $i/N$ by a continuous variable $x$ for large $N$. In the zero-temperature limit $\beta\rightarrow \infty$, this free energy reduces to
\begin{align}
f =  s(p-1) m^p-\int_{0}^1 dx \sqrt{(spm^{p-1})^2+\Gamma(x)^2},
\end{align}
which coincides with the free energy (\ref{eq:free_energy_zero_temp}) for the quantum model.  The finite-temperature phase diagram as depicted in Fig.~\ref{fig:SVMC} has qualitatively the same structure as the quantum counterpart, Fig.~\ref{fig:phase_diagram_finite_temp}, when the temperature is low. Therefore, as long as static properties in the large-$N$ and low-temperature limits are concerned, the SVMC model faithfully describes the behavior of the quantum system.

\section{Conclusions}
\label{sec:conclusion}
We have solved the ferromagnetic $p$-spin model with and without random longitudinal field under inhomogeneous driving of the transverse field.  The zero-temperature phase diagram for the case of ideal control of the transverse field, i.e., complete turning off of the field at each site, showed that the first-order transition that exists under homogeneous driving can be circumvented by inhomogeneous driving. Under nonideal circumstances, with a nonzero temperature or a nonzero value of the final transverse field, a new line of first-order transitions appears, which prevents us from avoiding a first-order transition. However, the new first-order transitions are weaker than the original one in the sense that the width of the free-energy barrier between local minima is smaller than in the original homogeneous case, which leads to an increase in the quantum tunneling rate. We therefore conclude that inhomogeneous driving of the transverse field has the potential to be at least quantitatively beneficial for a performance enhancement of quantum annealing.

It is not easy to understand why inhomogeneous driving mitigates the difficulties of first-order transitions. A phase transition is a phenomenon involving a large number of microscopic degrees of freedom simultaneously and cooperatively, resulting in a diverging correlation length in the case of a second-order transition.  The introduction of a spatiotemporal inhomogeneity of the driving field significantly reduces the number of microscopic degrees of freedom that are involved in the process of modification of the system properties at a given time, concurrently reducing the correlation length and modifying critical exponents, which may lead to the disappearance of transition as observed here. A theory based on a suppression of topological defects (Kibble-Zurek mechanism~\cite{Campo:2014aa}) via inhomogeneous driving in interacting spin systems that can be mapped onto a free fermionic system has been proposed in Ref.~\cite{Mohseni:2018aa}. Our mean-field approach complements this theory and leads to similar conclusions about the benefits of inhomogeneous driving.

Related is the problem of practical inhomogeneous driving protocols, e.g., the order in which spins are to be chosen to have the transverse field turned off.  In our mean-field-like model, all spins are equivalent in the cost function, and there is no specific way to choose a particular spin as the next target. Even spin $i=N$, which has its transverse field turned off immediately after the annealing process starts, points in the right direction thanks to the weak but non-negligible effective field from other spins, $-s(\sum_{i=1}^{N-1}\hat{\sigma}_i^z)^{p-1}/N$. This mechanism clearly comes from the uniform mean-field characteristics of the present problem and is not straightforward to generalize. The situation is nontrivial in general problems. Empirical protocols have been devised and tested on a physical quantum annealing device \cite{Lanting2017,Adame:2018aa}. Systematic theoretical guidelines remain to be established.

In practice, quantum annealing operates away from the adiabatic limit and is a dynamical process, and thus the static analysis in the present paper needs careful scrutiny before its conclusions are applied to practical situations. For example, though the static phase diagram is shared by the quantum model and the classical SVMC model in the ideal situation of zero temperature and complete turning off of the transverse field, the dynamical properties are expected to be quite different since quantum dynamics for large but finite-size systems allows tunneling through an energy barrier whereas there is no such mechanism classically at $T=0$. Nevertheless, dynamics is notoriously difficult to analyze since we should, in principle, solve the time-dependent Schr\"odinger equation directly, which is in general out of reach beyond small to moderate sizes.  It is encouraging in this respect that the static properties of the $p$-spin model are very much in accordance with the dynamical behavior in the case of reverse annealing.\footnote{Work in progress.} Further investigations of dynamics will shed more light on the relevance of the static analysis to physical quantum annealing, and are highly desired.

\acknowledgments
This work was partially funded by JSPS KAKENHI Grant No. 26287086. The research is based upon work partially supported by the Office of the Director of National Intelligence (ODNI), Intelligence Advanced Research Projects Activity (IARPA), via U.S. Army Research Office Contract No. W911NF-17-C-0050. The views and conclusions contained herein are those of the authors and should not be interpreted as necessarily representing the official policies or endorsements, either expressed or implied, of the ODNI, IARPA, or the U.S. Government. The U.S. Government is authorized to reproduce and distribute reprints for Governmental purposes notwithstanding any copyright annotation thereon.

\appendix
\begin{widetext}
\section{Derivation of the free energy}
\label{app:A}

In this appendix we derive the free energy Eq.~(\ref{eq:free_energy_zero_temp}) for the Hamiltonian in Eqs.~(\ref{eq:total_Hamiltonian})--(\ref{eq:driver_Hamiltonian}), following the standard procedure \cite{jorg2010energy,seki2012quantum}.

Using the Suzuki-Trotter decomposition, we can write the partition function as

\begin{align}
\label{eq:A_partition_function}
Z=\lim_{M\rightarrow \infty} Z_M =\lim_{M\rightarrow \infty} \Tr \left(e^{-(\beta/M)s\hat{H}_0}e^{-(\beta/M)\hat{V}}\right)^M =\lim_{M\rightarrow \infty} \Tr \left\{ \exp\left[\frac{\beta s N}{M}\left(\frac{1}{N}\sum_{i=1}^N\hat{\sigma}_i^z\right)^p\right] \exp \left[\frac{\beta}{M}\sum_{i=1}^N\Gamma_i\hat{\sigma}_i^x\right]\right\}^M,
\end{align}
where $\beta$ is the inverse temperature. For $M$ replicas, we insert the closure relation

\begin{align}
\hat{1}(\alpha) =\sum_{\{\sigma_i^z(\alpha)\}} \ket{\{\sigma_i^z(\alpha)\}}\bra{\{\sigma_{i}^z(\alpha)\}}\sum_{\{\sigma_{i}^x(\alpha)\}} \ket{\{\sigma_{i}^x(\alpha)\}}\bra{\{\sigma_{i}^x(\alpha)\}}
\quad (\alpha=1, 2, \cdots,M),
\end{align}
and obtain
\begin{align}
Z_M=\sum_{\{\sigma_i^z(\alpha)\}}\sum_{\{\sigma_i^x(\alpha)\}}\prod_{\alpha=1}^M \exp \left[\frac{\beta s N}{M} \left(\frac{1}{N}\sum_{i=1}^N \sigma_i^z(\alpha)\right)^p\right] \exp\left[\frac{\beta}{M} \sum_{i=1}^N \Gamma_i \sigma_i^x(\alpha) \right]  \prod_{i=1}^{N} \braket{\sigma_i^z(\alpha)|\sigma_i^x(\alpha)}\braket{\sigma_i^x(\alpha)|\sigma_i^z(\alpha+1)}.
\end{align}
Periodic boundary conditions are imposed by the trace operation, $\ket{\sigma_i^z(1)}=\ket{\sigma_i^z(M+1)}$.

To facilitate the calculations, we use the following relation
\begin{align}
\delta\left(N m(\alpha)-\sum_{i=1}^N \sigma_i^z(\alpha)\right)=\int d \tilde{m}(\alpha) \exp \left[-\tilde{m}(\alpha) \left(N m(\alpha) -\sum_{i=1}^N \sigma_i^z(\alpha)\right)\right]
\end{align}
and express the partition function as
\begin{align}
Z_M=&\sum_{\{\sigma_i^z(\alpha)\}}\sum_{\{\sigma_i^x(\alpha)\}}\prod_{\alpha=1}^M \int dm(\alpha) d\tilde{m}(\alpha) \exp\left[N\left(\frac{\beta s}{M} m(\alpha)^p -\tilde{m}(\alpha) m(\alpha)\right)\right]  \notag \\
&\times \exp\left[\sum_{i=1}^N \left(\tilde{m}(\alpha) \sigma_i^z(\alpha)+\frac{\beta}{M}\Gamma_i  \sigma_i^x(\alpha)\right)\right]\prod_{i=1}^{N} \braket{\sigma_i^z(\alpha)|\sigma_i^x(\alpha)}\braket{\sigma_i^x(\alpha)|\sigma_i^z(\alpha+1)} \notag \\
=&\int \prod_{\alpha=1}^M dm(\alpha) d\tilde{m}(\alpha) \exp\left[N\sum_{\alpha=1}^M\left(\frac{\beta s}{M} m(\alpha)^p -\tilde{m}(\alpha) m(\alpha)\right)\right]  \exp \left[\sum_{i=1}^N \ln \Tr \prod_{\alpha=1}^M \exp \left(\tilde{m}(\alpha) \hat{\sigma}^z\right) \exp \left(\frac{\beta}{M} \Gamma_i \hat{\sigma}^x\right)\right] \notag \\
=& \int \prod_{\alpha=1}^M dm(\alpha) d\tilde{m}(\alpha) \exp \left[-N\beta f_{N,M}\right].
\end{align}
For $N\gg 1$, the saddle-point condition for $\tilde{m}(\alpha)$ reads
\begin{align}
\tilde{m}(\alpha)=\frac{\beta s p}{M}m(\alpha)^{p-1}.
\end{align}
Then, the free energy becomes 
\begin{align}
f_{N,M}(\{m(\alpha)\})=&s(p-1)\frac{1}{M}\sum_{\alpha=1}^{M} m(\alpha)^p -\frac{1}{\beta N} \sum_{i=1}^N \ln \Tr \prod_{\alpha=1}^M \exp \left(\frac{\beta s p}{M} m(\alpha)^{p-1} \hat{\sigma}^z\right) \exp\left(\frac{\beta}{M}\Gamma_i  \hat{\sigma}^x\right),
\end{align}
\end{widetext}
We now use the static approximation $m=m(\alpha)$ for all $\alpha$. Taking the trace by the reverse operation of the Suzuki-Trotter decomposition for $M\rightarrow \infty$, we obtain
\begin{align}
f(m)=s(p-1) m^p -\frac{1}{\beta N}\sum_{i=1}^N \ln 2\cosh \beta \sqrt{(spm^{p-1})^2+\Gamma_i^2}.
\end{align}
The extremization condition of $f(m)$ leads to
\begin{align}
m=\frac{1}{N} \sum_{i=1}^N \frac{spm^{p-1}}{\sqrt{(spm^{p-1})^2+\Gamma_i^2}} \tanh \beta \sqrt{(spm^{p-1})^2+\Gamma_i^2}.
\end{align}
For $N\gg 1$, we rewrite $\Gamma_i$ with discrete valuable $i$ in terms of $\Gamma(x)$ with a continuous valuable $x\sim i/N$. Then the free energy and self-consistent equation reduce to
\begin{align}
&f(m)=s(p-1) m^p -\int_0^1 dx \ln 2\cosh \beta \sqrt{(spm^{p-1})^2+\Gamma(x)^2}, \\
&m=\int_0^1 dx \frac{spm^{p-1}}{\sqrt{(spm^{p-1})^2+\Gamma(x)^2}} \tanh \beta \sqrt{(spm^{p-1})^2+\Gamma(x)^2}.
\end{align}

\section{Semiclassical computations of the energy gap and the entanglement entropy }
\label{app:semiclassical}

We calculate in this appendix the energy gap in the limit $N\to\infty$ as quoted in Sec.~ \ref{sec:inhomogeneous_QA} and the entanglement entropy by the semiclassical method \cite{filippone2011quantum,seoane2012many}. The methods we employ are semiclassical since a large spin (for large $N$) behaves classically.

We divide the system into two subsystems $A$ and $B$, the former with $i=1,\cdots, N(1-\tau)$ and the latter for the rest of the sites. Note that according to our convention the transverse field is turned on in subsystem $A$ but is off in subsystem $B$. We further divide subsystem $A$ into two subsystems, $A_1$ with $i=1,\cdots, Nu(1-\tau)$, $A_2$ with $i=Nu(1-\tau)+1,\cdots, N(1-\tau)$, where $u$ is a parameter between $0$ and $1$. Our goal is to compute the energy gap and the entanglement entropy between the two subsystems $A_1$ and $A_2$ in the limit of large $N$.

To do so, we introduce two macroscopic spin operators as
\begin{align}
    \hat{S}_{A_1}^{z,x}&=\frac{1}{2}\sum_{i=1}^{Nu(1-\tau)} \hat{\sigma}_i^{z,x}, \\
    \hat{S}_{A_2}^{z,x}&=\frac{1}{2}\sum_{i=Nu(1-\tau)+1}^{N(1-\tau)} \hat{\sigma}_i^{z,x}, \\
    \hat{S}_{B}^{z,x}&=\frac{1}{2}\sum_{i=N(1-\tau)+1}^{N} \hat{\sigma}_i^{z,x}.
\end{align}
The Hamiltonian is then rewritten as
\begin{align}
    \hat{H}(s,\tau)= -sN \left\{\frac{2}{N}(\hat{S}_{A_1}^z+\hat{S}_{A_2}^z+\hat{S}_{B}^z)\right\}-2(\hat{S}_{A_1}^x+\hat{S}_{A_2}^x). \label{eq:appendix_H}
\end{align}
Rotating the spin operators around the $y$ axis by an angle $\theta$ as
\begin{align}
\left(
\begin{array}{c}
\hat{S}_{A_{1,2}}^x \\
\hat{S}_{A_{1,2}}^z
\end{array}
\right)
=
\left(
\begin{array}{cc}
\cos \theta &\sin \theta\\
-\sin \theta & \cos \theta
\end{array}
\right)
\left(
\begin{array}{c}
\hat{\tilde{S}}_{A_{1,2}}^x \\
\hat{\tilde{S}}_{A_{1,2}}^z
\end{array}
\right),
\end{align}
we employ the Holstein-Primakoff transformation to treat quantum corrections to the classical limit as
\begin{align}
    \hat{\tilde{S}}_{A_{1,2}}^z &= \frac{N_{1,2}}{2}-\hat{a}_{1,2}^{\dagger}\hat{a}_{1,2}, \\
    \hat{\tilde{S}}_{A_{1,2}}^+ &= (N_{1,2} - \hat{a}_{1,2}^{\dagger}\hat{a}_{1,2} )^{1/2}\hat{a}_{1,2} =(\hat{\tilde{S}}_{A_{1,2}}^-)^{\dagger}, \\
    \hat{S}_{B}^z &= \frac{N\tau}{2}-\hat{b}^{\dagger}\hat{b},\\
    \hat{S}_{B}^+ &= (N\tau - \hat{b}^{\dagger}\hat{b} )^{1/2}\hat{b} =(\hat{S}_{B}^-)^{\dagger},
\end{align}
where $N_1=Nu(1-\tau)$, $N_2=N(1-u)(1-\tau)$, and $\hat{a}_1$, $\hat{a}_2$, and $\hat{b}$ are bosonic annihilation operators.
Substituting these transformations into the Hamiltonian Eq.~(\ref{eq:appendix_H}) and expanding it to $\mathcal{O}(N^0)$ (the semiclassical limit), the Hamiltonian becomes
\begin{align}
    \hat{H}(s,\tau)= &Ne +\gamma +\delta (\hat{a}_1^{\dagger}\hat{a}_1+\hat{a}_2^{\dagger}\hat{a}_2) \notag \\
    & +\gamma \left[u\{(\hat{a}_1^{\dagger})^2+(\hat{a}_1)^2\}+(1-u)\{(\hat{a}_2^{\dagger})^2+(\hat{a}_2)^2\} \right. \notag \\
    &\left. +2\sqrt{u(1-u)}(\hat{a}_1^{\dagger}\hat{a}_2^{\dagger}+\hat{a}_1\hat{a}_2)\right] \notag \\
    & + \Delta_b \hat{b}^{\dagger} \hat{b},
\end{align}
where
\begin{align}
    e &= -s[\tau+(1-\tau)\cos \theta_0]^p-(1-\tau)\sin \theta_0, \\
    \gamma &= -\frac{1}{2}sp(p-1)(1-\tau) \sin^2 \theta_0 \{\tau+(1-\tau)\cos\theta_0 \}^{p-2}, \\
    \delta &= \Delta_b \cos\theta_0+2\sin \theta_0 +2\gamma,\\
    \Delta_b &= 2sp \{\tau+(1-\tau)\cos \theta_0 \}^{p-1}
\end{align}
with
\begin{align}
    \theta_0=\arg \min_{\theta}\left\{-s[\tau+(1-\tau)\cos \theta]^p-(1-\tau)\sin \theta\right\}.
\end{align}
To compute the energy gap and the entanglement entropy, we diagonalize the Hamiltonian using the Bogoliubov transformation as
\begin{align}
\hat{a}_1 &= \sqrt{u} \left\{\cosh \frac{\Theta}{2}\hat{\tilde{a}}_1+\sinh \frac{\Theta}{2}\hat{\tilde{a}}_1^{\dagger}\right\}+\sqrt{1-u}\,\hat{\tilde{a}}_2, \\
\hat{a}_2 &= \sqrt{1-u} \left\{\cosh \frac{\Theta}{2}\hat{\tilde{a}}_1+\sinh \frac{\Theta}{2}\hat{\tilde{a}}_1^{\dagger}\right\}-\sqrt{u}\,\hat{\tilde{a}}_2,
\end{align}
where 
\begin{align}
\tanh \Theta = -2\gamma/\delta=\epsilon ,
\end{align}
and $\hat{\tilde{a}}_1$ and $\hat{\tilde{a}}_2$ are new bosonic annihilation operators. The diagonalized Hamiltonian is given as
\begin{align}
    \hat{H}(s,\tau)=&Ne+\gamma+\frac{\delta}{2}(\sqrt{1-\epsilon^2}-1) \notag \\
    &+\Delta_{a_1} \hat{\tilde{a}}_1^{\dagger}\hat{\tilde{a}}_1+\Delta_{a_2} \hat{\tilde{a}}_2^{\dagger}\hat{\tilde{a}}_2+\Delta_{b} \hat{b}^{\dagger}\hat{b},
\end{align}
\begin{figure}
  \centering
  \includegraphics[width=0.49\linewidth]{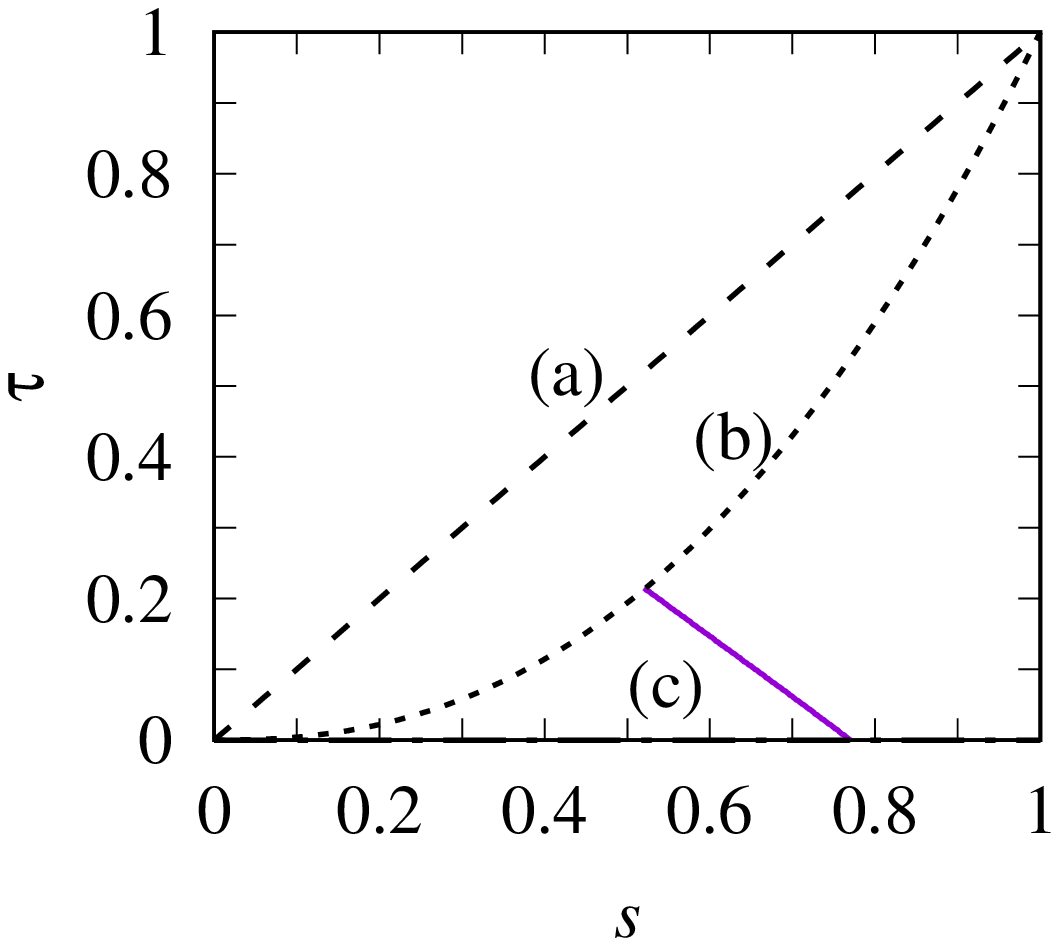}
  \includegraphics[width=0.49\linewidth]{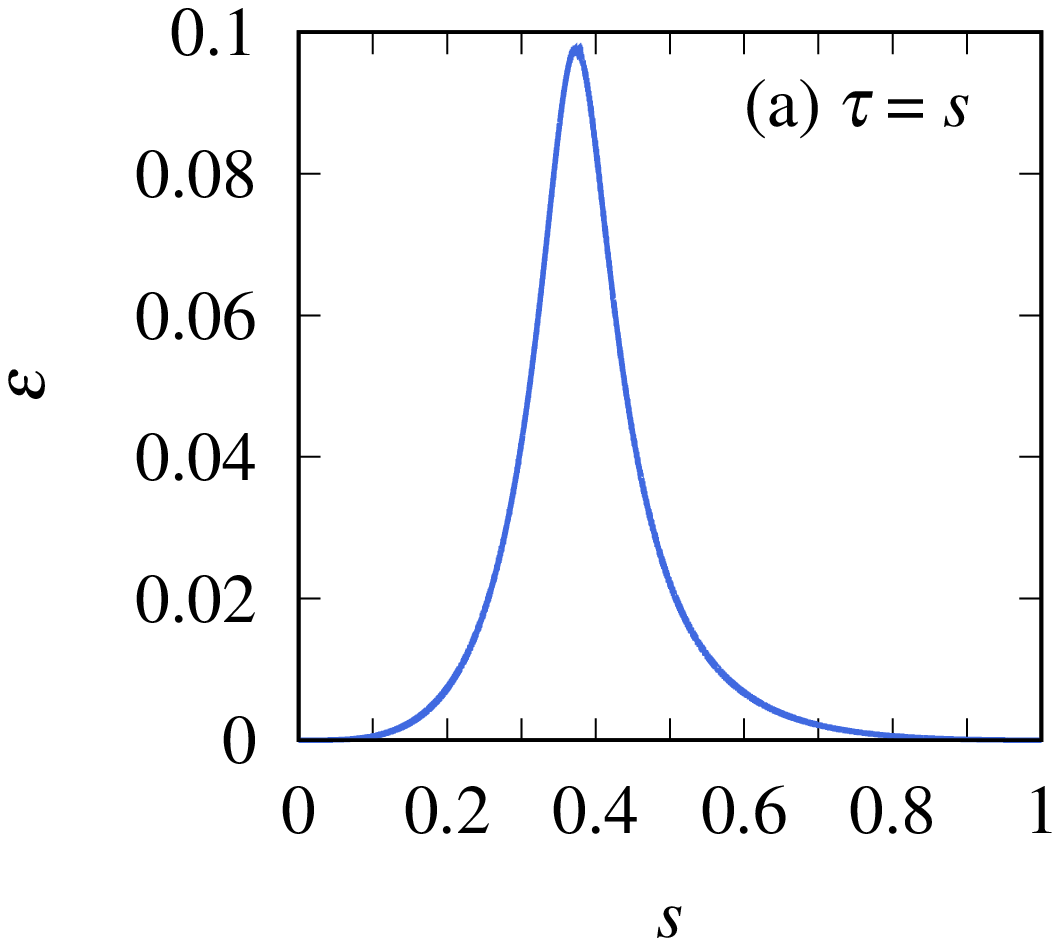} \\    \includegraphics[width=0.49\linewidth]{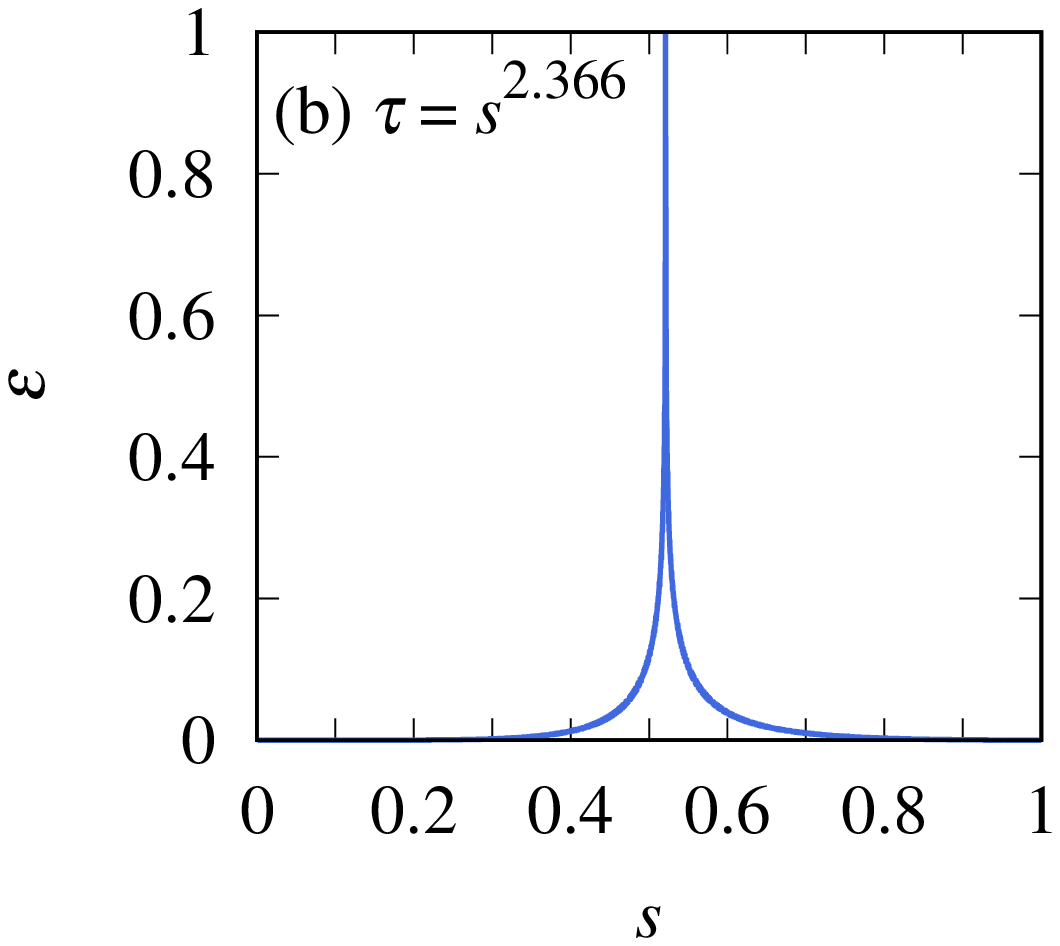}    \includegraphics[width=0.49\linewidth]{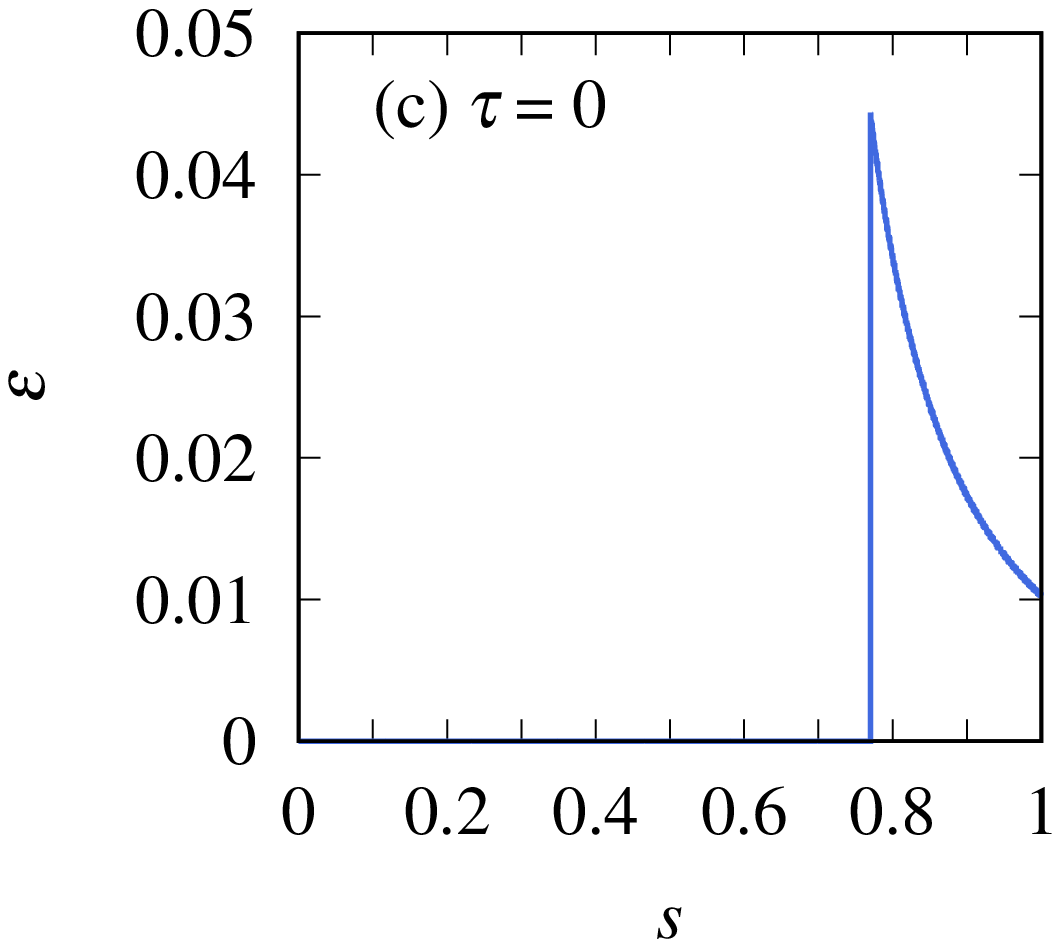}
  \caption{Left top panel is the phase diagram, where the solid line represents a line of first-order phase transitions, and three lines (a), (b), and (c) indicate paths with $\tau=s,\ s^{2.366}$ and $0$. Panels (a)--(c) show the entanglement entropy for the corresponding paths. In each case we set $p=3$ and $u=1/2$.}
   \label{fig:entanglement_entropy}
\end{figure}
where 
\begin{align}
    \Delta_{a_1} &= \delta\sqrt{1-\epsilon^2}, \\
    \Delta_{a_2} &= \delta.
\end{align}
Since $ \Delta_{a_2}\ge \Delta_{a_1}$, the minimum energy gap is the smaller of $\Delta_{a_1}$ and $\Delta_{b}$, i.e.,
\begin{align}
    \Delta ={\rm min}(\Delta_{a_1},\Delta_{b}).
\end{align}

The entanglement entropy between subsystems $A_1$ and $A_2$ is defined as $\mathcal{E}=-\Tr_{A_1} (\hat{\rho}_{A_1} \ln \hat{\rho}_{A_1})$, where $\hat{\rho}_{A_1}=\Tr_{A_2}\hat{\rho}_A$ is the density matrix of subsystem $A_1$ and $\hat{\rho}_A$ is the one for subsystem $A$. The technique for computing $\hat{\rho}_{A_1}$ is detailed in Ref.~\cite{filippone2011quantum} 
Using this method, the density matrix of subsystem $A_1$ is described as
\begin{align}
    \hat{\rho}_{A_1} = \frac{2}{\mu+1}\exp\left[-\ln \left(\frac{\mu+1}{\mu-1}\right)\hat{c}^{\dagger} \hat{c} \right],
\end{align}
where $\hat{c}^{\dagger}$ and $\hat{c}$ are bosonic creation and annihilation operators and
\begin{align}
    \mu &= \sqrt{[(1-u)+u\alpha][(1-u)+u/\alpha]}, \\
    \alpha &= \sqrt{(1-\epsilon)/(1+\epsilon)}.
\end{align}
The entanglement entropy $\mathcal{E}$ then becomes
\begin{align}
    \mathcal{E}=\frac{\mu+1}{2}\ln \frac{\mu+1}{2}-\frac{\mu-1}{2}\ln\frac{\mu-1}{2}.
\end{align}

Figure~\ref{fig:entanglement_entropy} shows the entanglement entropy $\mathcal{E}$ along three paths: (a) no crossing of the first-order transition line, (b) passing through the critical point, and (c) crossing the first-order transition line along the path corresponding to conventional QA ($\tau=0$). In the case (b) we can confirm that the entropy diverges continuously around the critical point. In contrast, for the case (c) a discontinuity exists at the transition point, a feature of a first-order phase transition~\cite{WuSarandyLidar:04}.

\bibliographystyle{apsrev4-1}

\end{document}